\documentclass[10pt]{article}

\usepackage{amsthm}
\usepackage{amsmath}
\usepackage{amssymb}
\usepackage{tikz}
\usepackage{graphicx}
\usepackage[footnotesize,hang]{caption}
\usepackage{subfig}
\usepackage{color}
\usepackage{rotating}
\usepackage{mathtools}
\usepackage{xcolor}
\usepackage{pifont}

\usepackage[margin=0.85in]{geometry}
\usetikzlibrary{decorations.pathmorphing}

\newcommand{\pdiff}[2]{\frac{\partial #1}{\partial #2}}
\renewcommand{\inf}{\infty}
\newcommand{\eps}{\epsilon}

\renewcommand{\i}{\mathrm{i}}
\newcommand{\e}{\mathrm{e}}

\renewcommand{\d}{\,\mathrm{d}}
\newcommand{\diff}[2]{\frac{\mathrm{d} #1}{\mathrm{d} #2}}

\def\Xint#1{\mathchoice
{\XXint\displaystyle\textstyle{#1}}%
{\XXint\textstyle\scriptstyle{#1}}%
{\XXint\scriptstyle\scriptscriptstyle{#1}}%
{\XXint\scriptscriptstyle\scriptscriptstyle{#1}}%
\!\int}
\def\XXint#1#2#3{{\setbox0=\hbox{$#1{#2#3}{\int}$}
\vcenter{\hbox{$#2#3$}}\kern-.5\wd0}}

\def\dashint{\Xint-}

\title{Exponential asymptotics for elastic and elastic-gravity waves on flow past submerged obstacles}
\author{Christopher J. Lustri$^1$\footnote{Electronic address: christopher.lustri@mq.edu.au}}
\date{%
$^1$School of Mathematical and Physical Sciences, 12 Wally's Walk, Macquarie University, New South Wales 2109, Australia\\
}                                   

\begin{document}
\maketitle

\abstract{Linearized flow past a submerged obstacle with an elastic sheet resting on the flow surface are studied \textcolor{black}{in the limit that the bending length is small compared to the obstacle depth}, in two and three dimensions. Gravitational effects are included in the two-dimensional geometry, but absent in the three-dimensional geometry; \textcolor{black}{the Froude number is chosen so that gravitational and elastic restoring forces are comparable in size}. In each of these problems, the waves are exponentially small in the asymptotic limit, and can be computed using exponential asymptotic methods. In the two-dimensional problem, flow past a submerged step is considered. It is found that the relative strength of the gravitational and elastic restoring forces produce two distinct classes of elastic sheet behaviour. In one parameter regime, constant-amplitude elastic waves and gravity waves extend indefinitely upstream and downstream from the obstacle. In the other parameter regime, all waves decay exponentially away from the obstacle. The equivalent nonlinear two-dimensional geometry is then studied; this asymptotic analysis predicts the existence of a third intermediate regime in which waves persist indefinitely only in one direction, depending on whether the submerged step rises or falls. In the three-dimensional geometry, it is predicted that the elastic waves extend ahead of the submerged source, decaying algebraically in space. The form of these elastic waves is computed, and validated by comparison with numerical computations of the elastic sheet behaviour. }

\section{Introduction}

Hydroelastic waves are waves that propagate due to flexural elasticity in a sheet or membrane which is resting on a fluid region. Early motivation for studying hydroelastic waves arose due to the use of elastic sheets as a convenient model for ice sheets floating on bodies of water \cite{Squire1, Squire2}. There have been a number of studies that considered moving bodies exerting pressure on an ice surface \cite{Davys1,dinvay2019fully,milinazzo1995mathematical,Parau2,Parau3,Schulkes1,Squire3,takizawa1985deflection}, as well as free and forced waves in ice sheets \cite{Guyenne3,guyenne2014forced}. Motivated by recently-proposed applications such as the use of piezoelectric membranes on the surface of a flow to harvest energy \cite{domino2018dispersion}, a number of recent laboratory experiments have studied the behaviour of hydroelastic waves on smaller scales \cite{akcabay2012hydroelastic, Ono1}. Many of these theoretical and experimental studies considered both gravitational and elastic restoring forces on the elastic sheet, producing wave behaviour known as flexural-gravity waves.

A number of studies of waves that form on elastic sheets resting on flow over submerged obstacles have been performed in linear regimes. In \cite{Sturova1, Sturova2, Sturova3}, the authors studied flow under finite or semi-infinite elastic plates past a submerged body in two dimensions using Green's function methods. The behaviour of a finite elastic plate was studied for flow over an obstacle in finite depth in \cite{Tkacheva1}. A linearized perturbation expression for the behaviour of an elastic sheet above a point source in infinite depth was obtained in \cite{Savin1}. Linearized geometries have also been studied using computational studies, such as the analysis in \cite{Shishmareva1}, which examined the elastic sheet strain caused by flow past a submerged dipole in a three dimensional channel using numerical Fourier methods.

Nonlinear models have been used to study solitary or periodic hydroelastic waves in both two and three dimensions. See, for example, the computational and asymptotic studies in \cite{Balmforth1, Forbes7, Forbes8, gao2016new, gao2018numerical, Guyenne1, guyenne2015asymptotic, Marchenko1, Milewski1, Parau2, Trichtchenko1, VandenBroeck6,wang2013two}. This is not an exhaustic list of research in this area; for a more comprehensive review, see \cite{Parau1}. Many of these studies use a nonlinear model for the elastic sheet deformation to express the surface wave behaviour in terms of an integrable equation such as the nonlinear Schr\''{o}dinger equation, which possesses soliton or periodic wave solutions. 

The behaviour of nonlinear flexural-gravity waves on flow past submerged obstacles in two dimensions was considered in \cite{stepanyants2021waves}, in which the authors derived a nonlinear Schr\''{o}dinger equation for weakly nonlinear perturbations above a submerged dipole, and \cite{Semenov1}, in which the authors studied fully nonlinear waves using a conformal mapping method developed in \cite{Forbes2,Forbes1,King1,King2,King3}. The problem was formulated by applying a conformal map taking the flow region, with an unknown free surface position, into a known domain. The resultant problem is expressed in terms of a boundary integral that can be solved numerically. 

Similar analyses have been performed on related geometries, including waves on internal flow interfaces under an elastic sheet \cite{wang2014numerical}, finite-depth shear flow under an elastic sheet \cite{wang2020progressive}, flow in a fluid separated by an internal elastic sheet \cite{puaruau2018solitary}, and flow contained between two elastic sheets \cite{blyth2011hydroelastic}. When studying nonlinear geometries, care must be taken in choosing an appropriate model for the elastic sheet; in \cite{Milewski2}, the authors investigated the effect of different elastic sheet models on nonlinear surface waves, and demonstrated that the choice of elastic sheet model can have a significant impact on the observed behaviour in nonlinear problems.  

Elastic waves have also been the subject of experimental studies \cite{domino2018dispersion, Ono1,akcabay2012hydroelastic}. Notably, \cite{Ono1} demonstrated that elastic waves in three dimensions exhibit similar qualitative behaviour to capillary waves, such as those computed in \cite{Lustri7}. The scaling regimes considered in this paper are comparable to laboratory setups such as \cite{Ono1}, which consider the waves that form on flexible elastic membranes suspended over an inviscid fluid. Motivated by this similarity, this paper aims to apply the exponential asymptotic techniques used in two-dimensional gravity-capillary waves in \cite{Trinh3,Trinh4} and three-dimensional capillary waves in \cite{Lustri7} to calculate the behaviour of hydroelastic waves.

\subsection{Paper outline}

We first study flexural-gravity waves in a linearized geometry generated by flow over a submerged step with small height, with elastic and gravity effects scaled by a small parameter, governing the rigidity of the elastic sheet and the Froude number (or the ratio between inertial and gravitational effects). This study is motived by previous analyses of gravity-capillary waves \cite{Trinh3,Trinh4}, which showed that the interaction between gravitational and capillary restoring forces produces a rich variety of wave behaviour compared to capillary waves in isolation. In \cite{Chapman6}, which studied capillary waves in two dimensions in the small surface tension limit, it was found using exponential asymptotic methods that capillary waves propagate with constant amplitude away from the disturbance. Similar methods were used to study gravity-capillary waves in linear \cite{Trinh3} and nonlinear geometries \cite{Trinh4}. Even in linear geometries, these studies demonstrated that there exist interactions between gravitational and capillary effects which affect the surface wave behaviour. Subsequent studies on elastic waves in the absence of gravity \cite{Lustri6} found that elastic wave behaviour in the absence of gravity has similar behaviour to the capillary waves studied in \cite{Chapman6}. 

This formulation will allow for direct comparison with the methods of \cite{Trinh3}, which demonstrated interaction effects between gravity and capillary waves in a similar scaling limit (small surface tension and small Froude number); they found that the wave behaviour changed depending on the parameter regime. In one regime, downstream gravity waves and upstream capillary waves propagate indefinitely without decay. In the other regime, the two waves decay exponentially in space away from the submerged obstacle. The purpose of this study is to determine whether a similar variety of wave behaviour is obtained by the inclusion of gravity effects into the elastic sheet geometry studied in \cite{Lustri6} -- we will identify a similar bifurcation structure in flexural-gravity waves. Finally, we will discuss the challenges required to extend this analysis to the nonlinear case, as in \cite{Trinh4}, and obtain some preliminary asymptotic results.

In the second part of this study, we investigate the behaviour of hydroelastic waves in three dimensions, and comparing these results with the three-dimensional capillary wave analysis in \cite{Lustri7}. We study waves on the surface of linearized flow past a submerged point source, in the limit that the elastic rigidity is small. This geometry requires a more complicated asymptotic analysis than the two-dimensional geometry, and we therefore consider only a regime in which gravitational effects may be neglected. This analysis provides a first step to a full three-dimensional flexural-gravity wave analysis; we will discuss the challenges involved in such an analysis in the conclusion of this paper. We will determine that the hydroelastic waves do possess important similarities with capillary waves from \cite{Lustri7}. The waves are absent immediately behind the obstacle, but appear as special curves on the free surface known as ``Stokes curves'' are crossed into the upstream region. 
 
In both of these problems, the surface waves are exponentially small \textcolor{black}{in the asymptotic parameter given by the ratio between the elastic bending length and the obstacle depth}. This makes the waves impossible to compute using classical asymptotic power series methods. Instead, we use exponential asymptotic methods to obtain the surface wave behaviour. Important early examples of exponential asymptotics being used to study free-surface flow over submerged obstacles are found in \cite{Chapman3,Chapman6}, which study exponentially small gravity and capillary waves respectively in regimes that make use of the full nonlinear dynamic boundary condition. These results were extended to linearized and nonlinear gravity-capillary waves in two dimensions \cite{Trinh3,Trinh4}, as well as gravity and capillary waves in linearized three-dimensional geometries \cite{Lustri2,Lustri4,Lustri7}. Recently, flexural waves through an elastic sheet in the absence of gravitational effects was studies in \cite{Lustri6}, using the full nonlinear boundary condition. The present study extends directly on this body of work, exploring elastic wave effects in more detail.

The layout of this paper is as follows. We begin by introducing several models that are used in existing literature to describe the behaviour of elastic sheets, and showing that these models are consistent in the linearized limit considered in the present study. We then briefly introduce the exponential asymptotic method that will be used to study flexural waves generated in elastic sheets. The remainder of the paper is divided into two parts: in the first part we calculate the behaviour flexural-gravity waves in a linearized two-dimensional geometry, and extend our analysis to make predictions about more complicated nonlinear geometries. In the second part, we calculate the behaviour of hydroelastic waves in a linearized three-dimensional geometry. The paper ends with conclusions and a discussion of the results, including an outline of the challenges expected in extending these results to nonlinear geometries, or introducing gravity into the three-dimensional geometry.

\subsection{Waves on Elastic Sheets}\label{S:elastic}

\subsubsection{Two-dimensional geometries}

\textcolor{black}{
Elastic sheets in two dimensions have often been studied using the Cosserat model, such that the pressure jump across the sheet is related to the curvature through
\begin{equation}
p = D \left(\kappa_{ss} + \frac{1}{2}\kappa^3\right),
\end{equation}
where $p$ is the pressure, $s$ is an arc length parameter, and $\kappa$ is the signed curvature, positive if the center of curvature lies in the fluid region. The flexural rigidity coefficient $D$ is given by $D = Eh^3/(12(1-\nu^2))$, where $E$ is the Young's modulus, $h$ is the plate thickness, and $\nu$ is the Poisson ratio. 
}

\textcolor{black}{
By linearizing two-dimensional flows and searching for waves of the form $\e^{\i(k x- \omega t)}$, it is possible to calculate the dispersion relation for flexural-gravity waves in the linear limit for flow on finite depth $L$; see, for example, the discussion in \cite{gao2018numerical}. The dispersion relation is given by 
\begin{equation}\label{2.dispersion_shallow}
\omega^2 = \left(g k + \frac{D k^5}{\rho}\right)\tanh(L k),
\end{equation}
where $\omega$ is the angular frequence, $k$ is the wavenumber, $D$ is the flexural rigidity, and $\rho$ is the fluid density. If the depth is taken to be large, so that $\tanh(L k) \to 1$, it is possible to determine a useful length scale for flexural--gravity waves. The phase velocity, $c = \omega/k$, in this regime is given by
\begin{equation}\label{2.dispersion}
c^2 = \frac{g}{k} + \frac{D k^3}{\rho}.
\end{equation}
The phase velocity has a minimum value $c_{\mathrm{min}}$ at $k = k_{\mathrm{crit}}$, where the group velocity and phase velocity are equal. These values are given by
\begin{equation}\label{2.critical}
\qquad k_{\mathrm{crit}} = \left(\frac{\rho g}{3D}\right)^{1/4},\qquad c_{\mathrm{min}} = 4 \left(\frac{Dg^3}{3\rho}\right)^{1/4}.
\end{equation}
If the waves are on a steady flow past an obstacle, waves can only form if the flow velocity $U$ exceeds $c_{\mathrm{min}}$. If $U > c_{\mathrm{min}}$, then the dispersion relation \eqref{2.dispersion} gives two solutions for the wavenumber. The larger solution ($k > k_{\mathrm{crit}}$) describes downstream gravity waves, and the smaller solution ($k < k_{\mathrm{crit}}$) describes upstream flexural waves. From the form of $k_\mathrm{crit}$ in \eqref{2.critical}, we see that the characteristic length scale at which the waves transition from the elastic regime to the gravity regime is proportional to $l_D = (D/(\rho g))^{1/4}$, where $l_D$ is known as the ``bending length''.
}

\textcolor{black}{The elastic sheets used in \cite{Ono1} have values of $D$ in the range $D \approx 6 \times 10^{-6}$ to $2 \times 10^{-8}$ N/m$^2$. If such an elastic sheet is suspended above water, such that $g \approx 10$ m/s$^2$ and $\rho \approx 10^3$ kg/m$^3$, the elastic bending length lies in the range $l_D \approx 0.5\times 10^{-3}$ to $1\times 10^{-3}$ m. Ice sheets in \cite{dimarco1993sea} were calculated to have values of $D$ lying in the range $D \approx 6\times 10^9$ to $9\times 10^9$ N/m$^2$. Using the same approximate values for $g$ and $\rho$, we find that the bending length of these ice sheets is $l_D \approx 28$ to 30 m. 
}

\textcolor{black}{
Section \ref{S.2D} of this study first considers linearized waves in a finite-depth channel containing a step with upstream depth $L$. The step height in the mapped potential plane after non-dimensionalization by $L$, denoted $w = \phi + \i \psi$, is given by $\delta$. We assume that $0 < \delta \ll 1$, producing a linearized regime. This assumption is equivalent to linearizing around small step height, or setting the ratio between the step height and channel depth -- which is $\mathcal{O}(\delta)$ as $\delta \to 0$, to be asymptotically small.
}

\textcolor{black}{After linearizing about the small step height, we then introduce a second small parameter into the problem, which describes the bending length to channel depth ratio, $l_D/L$. Using the physical parameters for elastic sheets defined above, if an elastic sheet is suspended above a fluid with depth of 1 cm, it has $l_D/L \approx 0.05$ to $0.2$. If an ice sheet is suspended above a channel of depth of 100 m, it has $l_D/L \approx 0.3$. Motivated by examples such as these, we are interested in studying problems in the asymptotic limit that $l_D/L$ is small.  
}

\textcolor{black}{In order to capture interactions between gravitational and elastic effects, we also set the Froude number, denoted $F$, to be small, with the relative scale chosen such that gravitational and elastic restoring effects are comparable in size. We define a new small parameter $\epsilon$ such that
\begin{equation}\label{2.froude}
F^2 = \frac{U}{g L} = \beta \epsilon, \qquad \left(\frac{l_D}{L}\right)^4 = \frac{D}{\rho g L^4} = \beta \tau \epsilon^4,
\end{equation}
where $\beta$ and $\tau$ are chosen so that we can adjust the relationship between the Froude number $F$ and the bending length to channel ratio $l_D/L$. This particular form for the small parameter $\epsilon$ is selected so that the subsequent analysis is analogous to \cite{Trinh3}, and the scaling of each quantity relative to $\epsilon$ is chosen so that both gravitational and elastic effects are described by our asymptotic results.
}

\textcolor{black}{
Note that the linearization step occurs before $\epsilon$ is defined. This implies that $0 < \delta \ll \epsilon \ll 1$ in the linearized problem, as we consider a full expansion of $\epsilon$ but only the leading-order equations for $\delta$. This regime allows us to establish the feasability of the method. In Section \ref{S:nonlinear2d}, we will study nonlinear flow over a step that is not an asymptotically small parameter. The problem formulation in this geometry will only contain the small parameter $\epsilon$, and the analysis will therefore be applicable to regimes where $0 < \epsilon \ll 1$. Much of the linearized analysis in Section \ref{S.2D} is performed in a manner that generalizes to the nonlinear problem.
}

\subsubsection{Three-dimensional geometries}

\textcolor{black}{A number of models have been used to describe elastic sheets resting on a fluid in three dimensions. The simplest linear elastic model for three dimensions is the biharmonic model. The pressure on the elastic sheet is derived using linearized beam theory, giving
\begin{equation}
p = D \Delta^2 \eta,\label{1.BH}
\end{equation}
where $\eta$ is the free surface height, and $\Delta$ is the biharmonic operator. This model has been used in \cite{Squire1,Squire3} to study the deformation of an elastic sheet resting on a fluid.  Nonlinear approaches have been considered in the literature such as a model based on the Cosserat theory of elastic shells, presented in \cite{guyenne2014forced,Milewski2, Trichtchenko1}. This model is given in \cite{Milewski2} by 
\begin{align}\nonumber
p = D\bigg(2\pdiff{}{x}[S(\eta_y \eta{xy} -& \eta_x\eta_{yy})] + 2\pdiff{}{y}[S(\eta_x \eta{xy} - \eta_y\eta_{xx})]  + \pdiff{^2}{x^2}[S(1+\eta_y^2)] - 2\pdiff{^2}{x\partial y}[S\eta_x\eta_y]\\
&+\pdiff{^2}{y^2}[S(1+\eta_x)^2] + \frac{5}{2}\pdiff{}{x}[S^2(1+|\nabla\eta|^2)^{3/2}\eta_x] + \frac{5}{2}\pdiff{}{y}[S^2(1+|\nabla\eta|^2)^{3/2}\eta_y] \bigg),\label{e.cosserat}
\end{align}
where
\begin{equation}
S = \frac{(1+\eta_x^2)\eta_{yy} + (1 + \eta_y^2)\eta_{xx} - 2\eta_x\eta_y\eta_{xy}}{(1+|\nabla\eta|^2)^{5/2}}.
\end{equation}
In our study of hydroelastic waves in three dimensions, we will apply the scaling $\eta = \delta \tilde{\eta}$, where $0 < \delta \ll 1$ and $\delta$ measures the strength of the submerged source. The Cosserat model in \eqref{e.cosserat} reduces to the biharmonic model in \eqref{1.BH} under this linearization. Hence, we will use the biharmonic model directly in our analysis of hydroelastic waves in three dimensions, noting that it describes behaviour produced by the commonly-used nonlinear Cosserat model in the linearized regime.}

\textcolor{black}{In the analysis of the three-dimensional problem, we will introduce a small parameter $\epsilon$ such that $\eps^3 = D/(\rho U^2 L^3)$, where $U$ is the upstream flow velocity and $L$ is a representative lengthscale in the problem. The asymptotic analysis performed on the linearized three-dimensional problem in the limit that $\epsilon$ is small. The regime in which the small $\epsilon$ analysis performed on the linearized geometry is valid is given by $0 < \delta \ll \epsilon \ll 1$. }

\subsection{Exponential Asymptotics}

In order to study the behaviour of hydroelastic waves, we will adapt the methodology of \cite{Trinh3} for the problem of two-dimensional flexural-gravity waves, and \cite{Lustri6} for the study of purely elastic waves in a three-dimensional setting. \textcolor{black}{These studies considered waves on a free surface due to gravity or capillary effects which were exponentially small in the small Froude number and surface tension limits respectively. In the present study, we will be performing an exponential asymptotic analysis in the limit that $\eps \to 0$, where $\eps$ is defined in \eqref{2.froude}. In this case, it governs both gravitational and elastic effects. The limit corresponds to geometries with small Froude number, as previously seen in the exponential asymptotic study of gravity waves in \cite{Chapman3}, and the ratio between the elastic bending length and the obstacle depth being small. These problems have the common property that they are singularly perturbed in the small parameter $\eps$, which appears in front of the leading derivative terms in the Bernoulli equation, shown after rescaling in \eqref{1:ndbc}. In singularly perturbed problems such as these, oscillatory behaviour in the solution such as surface waves have amplitude that is exponentially small in the asymptotic limit.}

Solutions to singularly-perturbed differential equations containing multiple exponential terms in the complex plane typically contain curves along which the behaviour of a subdominant exponential changes rapidly. These curves were first identified in \cite{Stokes1} and are known as ``Stokes curves''. This rapid change causes exponentially small oscillations to appear in the solution, such as the elastic waves considered in the present study. Asymptotic techniques have been developed for studying this exponentially small behaviour, collectively known as `exponential asymptotics'. A broad summary of these techniques may be found in \cite{Boyd1}  This investigation will apply the technique developed by \cite{Daalhuis1} and extended by \cite{Chapman1}, which utilises the rapid variation near Stokes curves in order to study exponentially small behaviour in solutions to ordinary and partial differential equations \cite{Chapman4}.

\textcolor{black}{The first step in this technique is to express the solution of the differential equation system as asymptotic power series in the independent variable $w$, such as
\begin{equation}\label{eq.intro1}
q(w) \sim \sum_{n=0}^{\inf} \eps^n q^{(n)}(w) \quad \mathrm{and} \quad \theta(w) \sim \sum_{n=0}^{\inf} \eps^n \theta^{(n)}(w)  \quad  \mathrm{as} \quad \eps \rightarrow 0,
\end{equation}
The series is typically divergent for singularly perturbed problems, and will not describe the behaviour of the surface waves, no matter how many series terms $q^{(n)}$ and $\theta^{(n)}$ are calculated. This is because the waves are exponentially small in the asymptotic limit, and therefore smaller than any algebraic power of $\eps$.}  Instead, we minimize the error of the divergent series approximation by truncating the series after a particular finite number of terms. To optimally truncate the series, we follow the heuristic described in \cite{Boyd1} and truncate the series after its smallest term. The optimal truncation point, denoted $N_{\mathrm{opt}}$. typically becomes large in the asymptotic limit, and hence identifying the asymptotic form of the ``late-order terms'' of the series (that is, the form of \textcolor{black}{$q^{(n)}$ and $\theta^{(n)}$} in the limit that $n \rightarrow \inf$) is sufficient to determine the smallest term in the series, and hence truncate the series optimally \cite{Chapman1}.

Dingle \cite{Dingle1} identified that successive terms in a divergent asymptotic series expansion generated by singularly perturbed equations like \eqref{1:ndbc} are typically obtained by repeated differentiation of earlier terms in the series. \textcolor{black}{This can be seen in the series recurrence relation \eqref{eq.1recur2}, in which the series terms $q^{(n-1)}$ and $\theta^{(n-4)}$ are differentiated once and four times respectively. This repeated differentiation will cause singularities in earlier terms to grown in strength as $n$ increases, and therefore persist into later terms.} As these singularities are repeatedly differentiated, the series terms typically diverge as the ratio between a factorial and the increasing power of a function $\chi$ which is zero at the singularity, ensuring that the late-order terms are also singular at this point. In \cite{Chapman1}, the authors proposed that the terms of a divergent asymptotic series generated in this fashion have asymptotic behaviour given by the sum of factorial-over-power ansatz expressions, each associated with a different early-order singularity. \textcolor{black}{For our two-dimensional problems, the proposed factorial-over-power behaviour is given by
\begin{equation}\label{eq.1ansatz}
q^{(n)} \sim \frac{Q\Gamma(n + \gamma)}{\chi^{n+\gamma}}\quad \mathrm{and} \quad \theta^{(n)} \sim \frac{\Theta \Gamma(n + \gamma)}{\chi^{n + \gamma}} \quad \mathrm{as} \quad n \to \infty,
\end{equation}
where $\Gamma$ is the gamma function defined in \cite{Abramowitz1},  $Q$, $\Theta$, $\gamma$ and $\chi$ are functions of $w$ that do not depend on $n$, and $\chi = 0$ at singularities of early series terms.  The global behaviour of the functions $Q$, $\Theta$, $\gamma$ and $\chi$ may be found by substituting this ansatz directly into the equations governing the terms of the asymptotic series, and matching to a rescaled local expansion of the solution in the neighbourhood of the singularity. }

The late-order term behaviour given in \eqref{eq.1ansatz} is related to applying a WKB (or Liouville-Green) ansatz of the form $A(w)\e^{-\chi(w)/\eps}$ to the equations for $q$ and $\theta$ linearized about the truncated expansion. It is clear from this expression that $\chi$, or the ``singulant'', determines the scaling of the exponentially small terms. The rapid change in exponentially small behaviour, or ``Stokes switching'', occurs across curves where the switching exponential is maximally subdominant compared to the leading-order behaviour \cite{Dingle1}. These curves satisfy the condition that the singulant is purely real and positive, giving the following condition that may be used to determine the possible location of Stokes lines:
\begin{equation}\label{1.StokesCond}
\mathrm{Re}(\chi) > 0,\qquad \mathrm{Im}(\chi) = 0.
\end{equation}
Asymptotic solutions also contain important curves known as anti-Stokes lines. These are curves divide the complex plane into regions in which a particular exponential contribution is asymptotically small, and regions in which the exponential contribution is asymptotically large. From the WKB ansatz of the exponential contribution, it can be seen that anti-Stokes curves satisfy
\begin{equation}\label{1:AntiStokesCond}
\mathrm{Re}(\chi) = 0.
\end{equation}
\textcolor{black}{Truncating the infinite series \eqref{eq.intro1} optimally after $N$ terms gives
\begin{equation}\label{eq.intro1_trunc}
q(w) = \sum_{n=0}^{N_{\mathrm{opt}}-1} \eps^n q^{(n)}(w) + R_N(w) \quad \mathrm{and} \quad  \theta(w) = \sum_{n=0}^{N_{\mathrm{opt}}-1} \eps^n \theta^{(n)}(w) + S_N(w)  \quad  \mathrm{as} \quad \eps \rightarrow 0,
\end{equation}
where  $R_N$ and $S_N$ are the exponentially small remainder terms for $q$ and $\theta$ obtained after truncation.} Note that this expression is an equality, rather than an asymptotic relation. $R_N$ and $S_N$ therefore represent the difference between the true solution and the optimally truncated series.

The final step of the method described in \cite{Daalhuis1} requires substituting the truncated series expression back into the original problem to produce an equation for the remainder term. This remainder equation is then solved in the neighbourhood of Stokes curves, which are found using the condition in \eqref{1.StokesCond}\footnote{Condition \eqref{1.StokesCond} is not strictly required for this step, as the location of the Stokes curves can be obtained directly from the remainder equations using late-order terms. We will use this condition, as it allows for the Stokes curves to be identified once $\chi$ has been calculated, rather than later in the analysis.}. \textcolor{black}{This analysis shows that the exponentially small remainder that switches across the Stokes line generated by the truncated divergent series \eqref{eq.intro1_trunc} generally takes the form
\begin{equation}\label{eq.RN}
R_N \sim \mathcal{S} Q\e^{-\chi/\eps} \quad \mathrm{and} \quad S_N \sim \mathcal{S} \Theta\e^{-\chi/\eps} \quad \mathrm{as} \quad \eps \rightarrow 0,
\end{equation}
where $\mathcal{S}$ is a function of $w$ that is essentially constant away from the Stokes curve, but varies rapidly in the neighbourhood of the Stokes curve.} This emphasises the important role played by the singulant in determining the behaviour of the oscillations. Importantly, if $\chi'$ is purely imaginary, these terms correspond to exponentially small oscillations \textcolor{black}{as $\eps \to 0$} that do not decay exponentially in space. \textcolor{black}{If $Q$ and $\Theta$} do not decay, as is the case for the oscillations in the present study, then these terms produce a train of waves with constant amplitude.

\section{Two-Dimensional Elastic-Gravity Waves}\label{S.2D}

\subsection{Formulation}

\begin{figure}
\centering
\includegraphics{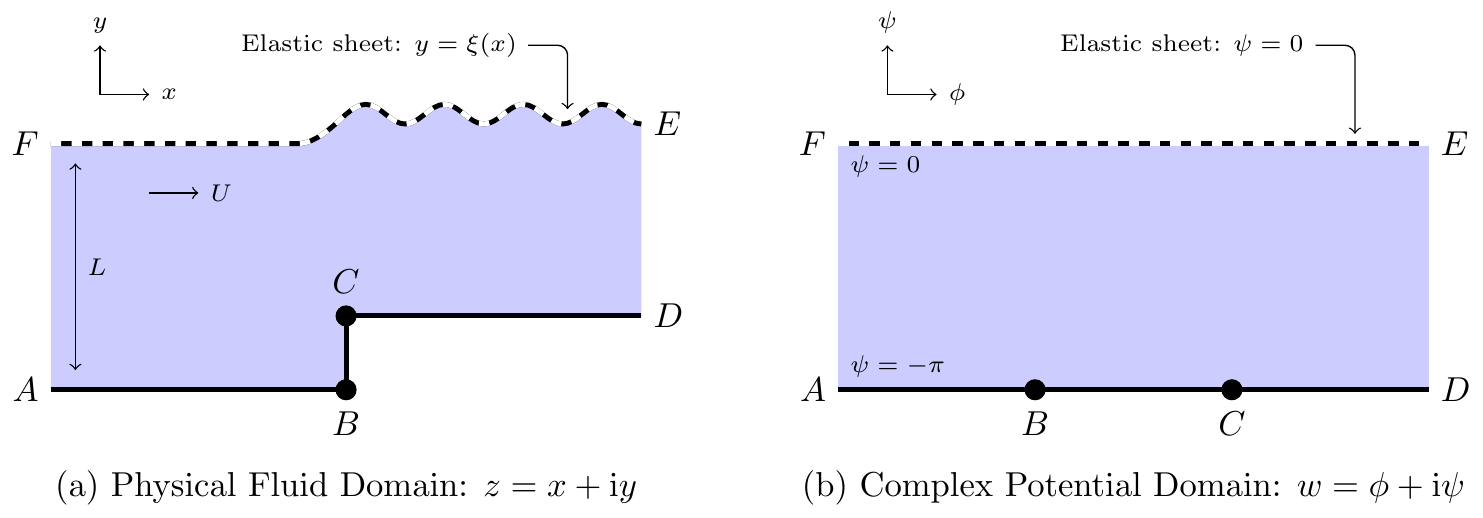}

\caption{Mapping of a fluid domain underneath an elastic sheet to a fixed known region of the complex potential plane. The elastic sheet is shown as a dashed line, and the rigid base is shown as an unbroken line. Steady flow follows streamlines, which are curves in the potential plane with constant $\psi$. The free surface maps to the top streamline, typically labelled $\psi = 0$, while the lower boundary typically maps to $\psi = -\pi$; consequently the flow region is known completely. The fluid velocity is singular at the points labelled $B$ and $C$, shown as black circles.}\label{Fig1}

\end{figure}
We consider a \textcolor{black}{two}-dimensional incompressible, irrotational, inviscid flow through a channel of finite depth over a submerged step. The upstream channel depth is given by $L$, and the upstream flow velocity is given by $U$. An elastic sheet with flexural rigidity $D$ rests on the surface of the flow.  The position of the elastic sheet is denoted as $\xi(x)$. A schematic of this flow behaviour is shown in Figure \ref{Fig1}. We now non-dimensionalise the lengths of the system by the upstream depth $L$, and the velocities by the upstream flow velocity $U$.

The fluid potential satisfies Laplace's equation
\begin{equation}
\nabla^2 \phi = 0.
\end{equation}
As the flow is steady, we apply a kinematic boundary condition on all boundaries,
\begin{equation}
\pdiff{\phi}{n} = 0,
\end{equation}
where $n$ is the unit normal direction. \textcolor{black}{On the free surface, we have the dynamic boundary condition, obtained from the Bernoulli equation,
\begin{equation}\label{2.physbernoulli0}
\frac{F^2}{2}(|\nabla \phi^2| - 1) +  y +  \frac{D}{\rho g L^4} \left(\kappa_{ss} \textcolor{black}{+ \frac{1}{2}\kappa^3}\right) = 0\quad \mathrm{on}  \quad y = \xi(x),
\end{equation}
where $\kappa$ is the curvature, defined to be positive if the center of curvature lies within the fluid, $s$ is the arc length along the surface after non-dimensionalisation, $\rho$ is the density of the fluid, $g$ is acceleration due to gravity, $F$ is the Froude number defined in \eqref{2.froude}, and $D$ is the flexural rigidity of the plate. The upstream flow is uniform with velocity $U$, such that
\begin{equation}\label{2.upstream}
(\phi_x, \phi_y) \to (1,0) \quad \mathrm{as} \quad x \to -\infty.
\end{equation}
}
Using the quantities defined in \eqref{2.froude}, we may rewrite \eqref{2.physbernoulli0} as
\begin{equation}\label{1:ndbc}
\frac{\beta \eps}{2}(|\nabla \phi|^2 - 1) + y + \beta \tau \eps^4  \left(\kappa_{ss}  \textcolor{black}{+ \frac{1}{2}\kappa^3}\right)  = 0\quad \mathrm{on}  \quad y = \xi(x),
\end{equation}
where $\epsilon$ is a small parameter, and $\beta$ and $\tau$ determine the ratio between the Froude number $F$ and the elastic length ratio $l_D/L$. We note that the $\tau = 0$ problem corresponds to pure gravity waves, studied in \cite{Chapman3}, while $\beta \to 0$ and $\tau = 1/\beta$ corresponds to pure elastic waves, studied in \cite{Lustri6}. In Section \ref{S:EA1}, we will find our results to be consistent with these prior studies. 

Differentiating \eqref{1:ndbc} with respect to $s$ gives
\begin{equation}\label{eq.Bern_diff}
\beta\eps q \diff{q}{s} + \diff{y}{s} + \beta \tau \eps^4 \left(\kappa_{sss}  \textcolor{black}{+ \frac{3}{2}\kappa_s \kappa^2}\right)  = 0\quad \mathrm{on}  \quad y = \xi(x).
\end{equation}
We define a complex potential $w = \phi + \i \psi$, where $\phi$ is the fluid potential, and $\psi$ is the streamfunction. This maps the fluid region to an infinite strip bounded by $\psi = -\pi$ and $\psi = 0$.  Noting that 
\begin{equation}
\kappa = \diff{\theta}{s},\qquad \diff{y}{s} = \sin\theta,  \qquad \diff{}{s} = q \diff{}{\phi},
\end{equation}
we write the Bernoulli condition \eqref{eq.Bern_diff} in terms of $\phi$.  This gives
\begin{align}
\nonumber \beta\eps q^2 \diff{q}{\phi} + \sin\theta +& \beta \tau \eps^4 \Bigg[ q \left(\diff{q}{\phi}\right)^3 \diff{\theta}{\phi} + 4q^2\diff{q}{\phi} \diff{\theta}{\phi}\diff{^2 q}{\phi^2}+ 7q \left(\diff{q}{\phi}\right)^2\diff{^2\theta}{\phi^2} \\
&+ 4 q^3\diff{^2 q}{\phi^2}\diff{^2\theta}{\phi^2} + q^3\diff{\theta}{\phi}\diff{^3\theta}{\phi^3} + q^4 \diff{^4\theta}{\phi^4} \textcolor{black}{+ \frac{3}{2}q^3\diff{q}{\phi}\left(\diff{\theta}{\phi}\right)^3 + \frac{3}{2}q^4\left(\diff{\theta}{\phi}\right)^2\diff{^2\theta}{\phi}}\Bigg] = 0.\label{2.physbern}
\end{align}
We also define the complex velocity $\d w/\d z = u - \i v$, written as $q \e^{-\i\theta}$. In this formulation, $q$ is the flow velocity at a point, and $\theta$ is the angle the streamlines make with the horizontal axis. Analytically continuing \eqref{2.physbern} allows us to replace $\phi$ with the complex potential $w$ in \eqref{2.physbern} to obtain the analytically-continued free surface condition
\begin{align}
\nonumber \beta\eps q^2 \diff{q}{w} + \sin\theta + \beta& \tau \eps^4 \Bigg[ q \left(\diff{q}{w}\right)^3 \diff{\theta}{w} + 4q^2\diff{q}{w} \diff{\theta}{w}\diff{^2 q}{w^2}+ 7q \left(\diff{q}{w}\right)^2\diff{^2\theta}{w^2} \\
&+ 4 q^3\diff{^2 q}{w^2}\diff{^2\theta}{w^2} + q^3\diff{\theta}{w}\diff{^3\theta}{w^3} + q^4 \diff{^4\theta}{w^4} \textcolor{black}{+ \frac{3}{2}q^3\diff{q}{w}\left(\diff{\theta}{w}\right)^3 + \frac{3}{2}q^4\left(\diff{\theta}{w}\right)^2\diff{^2\theta}{w}}\Bigg] = 0.\label{2.potentialbern}
\end{align}

\begin{figure}
\centering
\includegraphics{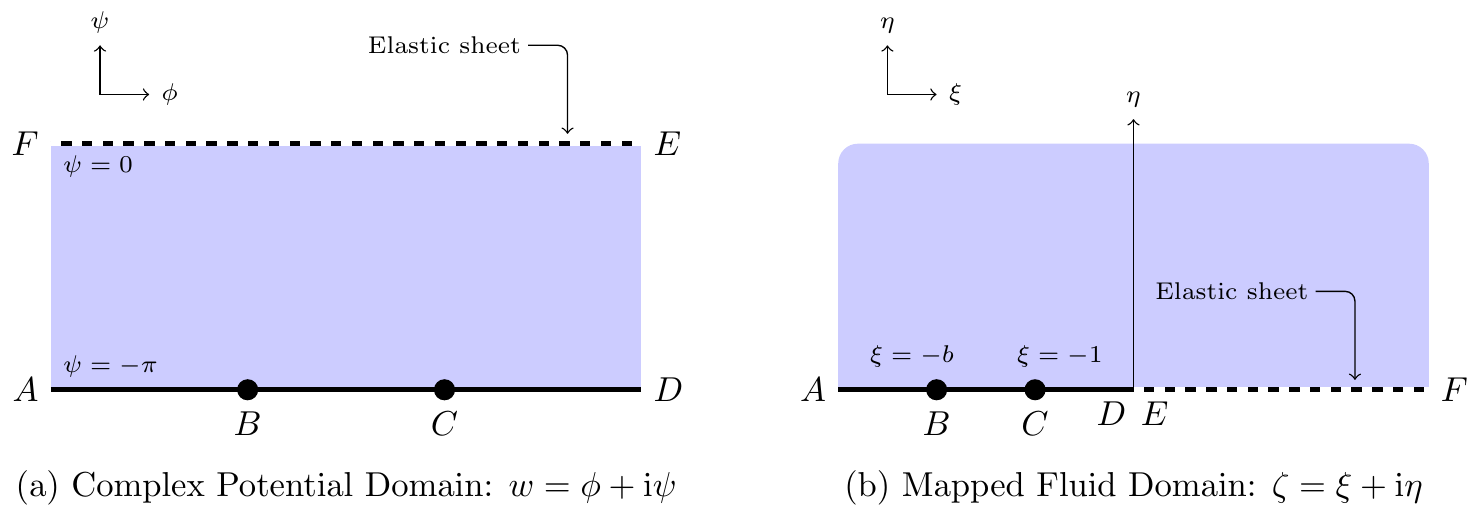}

\caption{This schematic illustrates the effect of the mapping $w \mapsto \zeta$ between the fluid potential domain, shown in (a), and the mapped domain, shown in (b). The mapping takes the fluid region to the entire upper half mapped plane. The elastic sheet $\psi = 0$ maps to the line $\xi > 0$, and the submerged boundary $\psi = -\pi$ maps to the line $\xi < 0$. The elastic sheet is shown as a dashed line, and the rigid base is shown as an unbroken line. The singularities map to points which will be labelled $\zeta = -b$ and $\zeta = -1$.}\label{Fig4}
\end{figure}

We apply a conformal map $\zeta = \e^{-w}$ in order to map the fluid region from a strip in the complex potential plane to the upper half $\zeta$-plane. We also define $\zeta = \xi + \i \eta$, where $\xi$ and $\eta$ are real quantities. Notably, the free surface maps to $\xi > 0$ and the base of the flow region maps to $\xi < 0$. In the mapped plane, we can apply Cauchy's theorem to obtain
\begin{equation}
\log q = - \frac{1}{\pi} \dashint_{-\infty}^{\infty} \frac{\theta(\xi')}{\xi' - \xi}\d\xi'.
\end{equation}
Analytically continuing this expression into the upper half-plane gives
\begin{equation}
q - \i \theta = -\frac{1}{\pi}\int_{-\infty}^{\infty} \frac{\theta(\xi')}{\xi' - \zeta}\d\xi'.
\end{equation}
We will define the base of the flow as a step, where $\theta = \pi/2$ for $-(1+\delta) < \zeta < -1$, and $\theta = 0$ for $\zeta < -(1+\delta)$ and $-1 < \zeta < 0$. Hence,
\begin{equation}
q - \i \theta = \frac{1}{2}\log\left(\frac{\zeta+b}{\zeta+1}\right)-\frac{1}{\pi}\int_{0}^{\infty} \frac{\theta(\xi')}{\xi' - \zeta}\d\xi'.\label{2.zetacauchy}
\end{equation}
We note that the strip in the complex potential plane may also be analytically-continued into the lower half $\zeta$-plane, which will produce complex conjugate behaviour. The full behaviour of the elastic sheet can be obtained by taking the sum of both the upper and lower half plane contributions. We will not perform the lower half plane calculations explicitly, but will instead add the appropriate complex conjugate contribution to the results of the upper half plane analysis.

This completes the governing equations. We treat \eqref{2.potentialbern} and \eqref{2.zetacauchy} as the equations governing the analytically-continued free surface. We will subsequently express the Bernoulli equation in terms of the mapped variable $\zeta$, but we will hold off until after the linearization step.

\subsection{Linearization}

We can linearize around the free stream for small step height. We set $b = 1 + \delta$ (assuming $0 < \delta \ll \eps$) and set $q = 1 + \delta \hat{q}$ and $\theta = \delta \hat{\theta}$. The linearized fluid equation is now given by
\begin{equation}
\hat{q} - \i \hat{\theta} = \frac{1}{2(\zeta+1)}-\frac{1}{\pi}\int_{0}^{\infty} \frac{\hat{\theta}(\xi')}{\xi' - \zeta}\d\xi'.
\end{equation}
The linearized Bernoulli equation is given by
\begin{equation}
\beta\eps \diff{\hat{q}}{w} + \hat{\theta} + \beta\tau\eps^4 \diff{^4\hat{\theta}}{w^4} = 0.
\end{equation}
We could apply the mapping $\zeta = \e^{-w}$ to the Bernoulli equation, but the analysis is more straightforward in the complex potential plane. We have now fixed the problem so that the boundary follows a known curve ($\zeta > 0$ in the mapped plane, $\psi = 0$ in the complex potential plane). The linearization step is not necessary; we could apply this exponential asymptotic analysis to the fully nonlinear problem. In this case, the existence of both upstream and downstream waves on the free boundary mean that it is difficult to verify the results computationally. Progress on studying these systems has been made in the context of gravity-capillary waves \cite{Jamshidi1}, but this remains a challenging numerical problem. We will briefly discuss the asymptotics of nonlinear geometries in Section \ref{S:nonlinear2d}.

\subsection{Exponential Asymptotics}\label{S:EA1}

We write the series expression in the limit that $\eps \to 0$,
\begin{equation}
\hat{q} \sim \sum_{n=0}^{\infty}\eps^{n}q^{(n)},\qquad \hat{\theta} \sim \sum_{n=0}^{\infty}\eps^{n}\theta^{(n)}.
\end{equation}
Note that including both gravity and elastic waves means that the form of the late-order terms requires powers of $\eps$ rather than $\eps^3$, unlike \cite{Lustri6}. \textcolor{black}{Note that retaining the full power series in $\epsilon$ in a system that has been linearized in $\delta$ implies that we are considering the regime $0 < \delta \ll \epsilon \ll 1$.}

The leading-order behaviour of the flow on the complex free surface is found by direct substitution, giving
\begin{equation}\label{2.q0lin}
q^{(0)} = \frac{1}{2(\zeta+1)} = \frac{1}{2(\e^{-w}+1)},\qquad \theta^{(0)} = 0.
\end{equation}
The leading-order behaviour is singular at $w = \pm (2M+1) \pi \i$, for $M \in \mathbb{Z}$. The singularities that matter are located at $M = \pm 1$. We will concentrate on the contributions due to the singularity at $w =\pi \i$, adding the corresponding contributions afterwards. At higher orders, we obtain recurrence expressions for the complexified free surface
\begin{align}\label{eq.1recur1}
q^{(n)} - \i \theta^{(n)} = -\frac{1}{\pi}\int_{0}^{\infty} \frac{\theta^{(n)}(\xi')}{\xi' - \zeta}\d\xi',\\
\beta  \diff{q^{(n-1)}}{w} + \theta^{(n)} + \beta \tau  \diff{^4\theta^{(n-4)}}{w^4}= 0.\label{eq.1recur2}
\end{align}
We are most interested in the form of the late-order terms, so we apply the late-order ansatz as $n \to \infty$ from \eqref{eq.1ansatz}. In the first equation from \eqref{eq.1recur1}, we neglect the integral expression, as it must be exponentially subdominant to the remaining terms in the expression. This simplification was used in \cite{Chapman6,Chapman3}, and discussed in detail for the case of gravity waves past a ship in \cite{Trinh1}. A similar justification can be made here. At leading order as $n \to \infty$, we find $Q = \i \Theta$, and
\begin{equation}\label{eq.singulant1}
1 - \i\beta  \diff{\chi}{w} + \beta \tau \left(\diff{\chi}{w}\right)^4 = 0.
\end{equation}
We can determine the form of the exponentially small contributions by solving the singulant equation \eqref{eq.singulant1}. We recall that $\chi = 0$ at $w = \pi \i$. This expression has four solutions, of the form
\begin{equation}
\chi =  k_j (w + \i \pi),\label{1:chi}
\end{equation}
where $k_j$ for $j = 1,\ldots, 4$ depends on $\beta$ and $\tau$, but not $\zeta$. We denote the specific singulants as $\chi_j$ for $j = 1,\ldots,4$.

Continuing to the next order, corresponding to $\mathcal{O}(q^{(n-1)})$ as $n \to \infty$, gives $Q$ and $\Theta$ constant. To clearly denote this, we write $\Theta = \Lambda$ and $Q = \i\Lambda$, where $\Lambda$ is constant in $w$, although it does depend on $\beta$ and $\tau$. This term must be determined by comparing the late-order terms with the inner problem in the neighbourhood of the singularity at $\zeta = -\i\pi$. We perform this inner analysis in Appendix \ref{S:2Dinner}, and find that the prefactors are given by 
\begin{equation}
\Lambda_j =  \frac{k_j^8(k_j A_5 + k_j^2 A_6 + k_j^3 A_7 -\tfrac{1}{\beta\tau} A_8 )}{ \tfrac{2}{\beta\tau} - 6 k_j^4},
\end{equation}
where $A_j = \i^j \beta^{j-3}(\beta^3 + (4-j)\tau)$ for $j = 5, \ldots, 8$. The index choice is related to the analysis in Appendix \ref{S:2Dinner}.

Knowing that $\Theta$ and $Q$ are constants, we can determine the value of $\gamma$ that is required for the late-order terms to be consistent with the leading-order behaviour near the singularity. The singularity in the leading order has strength one, and this will increase by one at each iteration. Hence, the strength of the singularity in the series term $q^{(n)}$ will be $n+1$, indicating that $\gamma = 1$. Consequently, we have fully determined the late-order asymptotic series terms \eqref{eq.1ansatz}.

Using the methods of \cite{Chapman1,Daalhuis1}, shown in detail in \cite{Trinh3,Trinh4}, we may determine the behaviour of the surface waves that correspond with each of the four solutions to \eqref{1:chi}. This analysis is presented in Appendix \ref{S:2Dsmoothing}, and gives the exponentially small wave contribution from $\chi_j$, which we denote as $\theta_{\mathrm{exp},j}$, as
\begin{equation}
\theta_{\mathrm{exp},j} \sim 2\, \mathrm{Re}\left[\left(\frac{1}{1 - 4 \tau \i k_j^3}\right)\frac{2\pi k_j \Lambda_j}{\beta\eps}\e^{-k_j(w+\i\pi)/\eps}\right],\label{2d:expsm}
\end{equation}
with a similar expression for $q_{\mathrm{exp},j}$. The real part is obtained by taking the sum of the upper and lower half $\zeta$-plane contributions, which are complex conjugate values. We have therefore calculated the form of the waves, and can determine the regions in which they are present by studying the Stokes phenomenon in the system.

\begin{figure}
\centering
\subfloat[Values of $k_j$ for $\tau = 1$.]{
\includegraphics{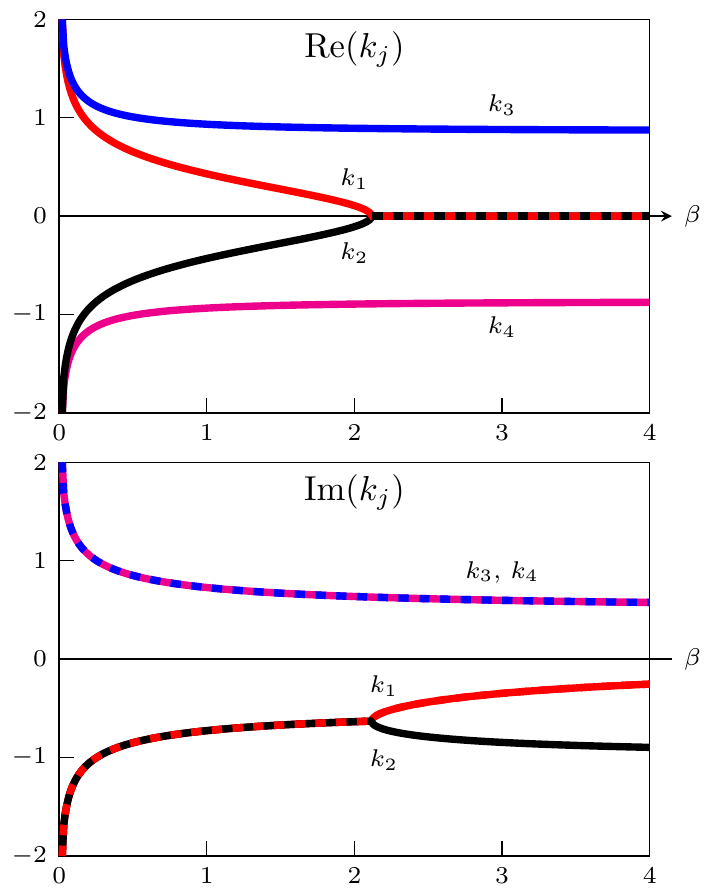}
}
\subfloat[Values of $k_j$ for $\beta = 1$.]{
\includegraphics{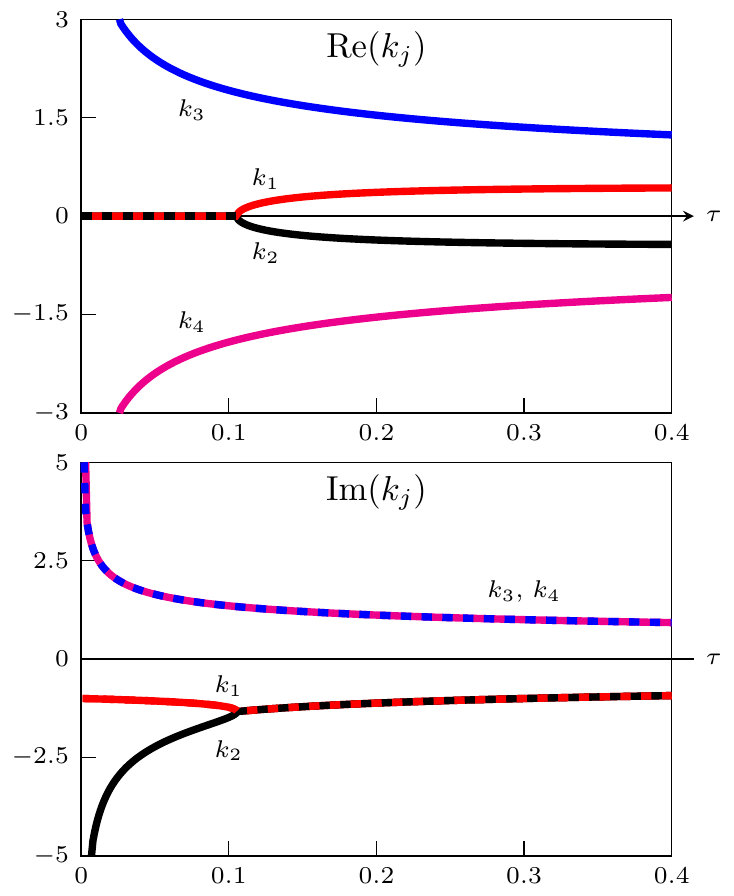}
}

\caption{Real and imaginary component of $k_1$ (black), $k_2$ (red), $k_3$ (blue), and $k_4$ (magenta) for (a) $\tau = 1$, and (b) $\beta = 1$. Dashed lines indicate multiple solutions $k_j$ taking identical value. In (a), there is a critical value of $\beta$ above which $\mathrm{Re}(k_{1,2}) = 0$. In (b), there is a critical value of $\tau$ below which $\mathrm{Re}(k_{1,2}) = 0$.}\label{fig:b-tau1}
\end{figure}

\begin{figure}[tb]
\centering
\includegraphics{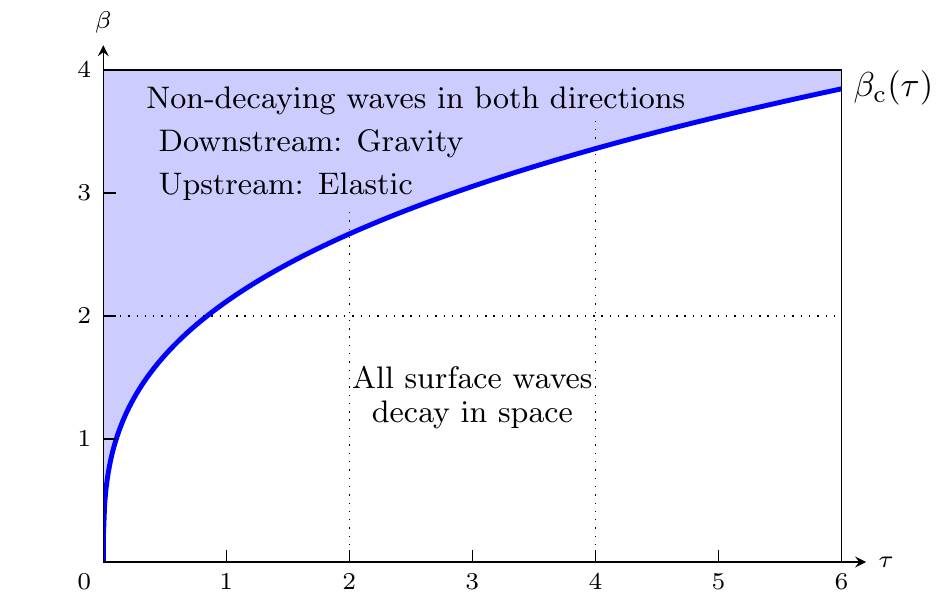}

\caption{Illustration of the wave behaviour as $\beta$ and $\tau$ are varied. In the unshaded region, corresponding to $\beta < \beta_{\mathrm{c}}$, all of the surface waves decay spatially away from the obstacle. In the shaded region, corresponding to $\beta > \beta_{\mathrm{c}}$, the surface behaviour contains non-decaying waves in the downstream and upstream directions, driven by gravitational and elastic restoring forces respectively.}\label{fig:b-tau2}
\end{figure}

From the value of $\chi$ in \eqref{1:chi} and the form of the exponential oscillations \eqref{eq.RN}, it is apparent that non-decaying wave behaviour can only exist if $k_j$ takes a purely imaginary value. In Figure \ref{fig:b-tau1} (a), we illustrate the solutions for $\tau = 1$ over a range of $\beta$, while in Figure \ref{fig:b-tau1} (b) we illustrate the solutions for $\beta = 1$ over a range of $\tau$. For fixed $\tau$, there exists some critical $\beta$, denoted $\beta_{\mathrm{c}}(\tau)$ such that two values of $k_j$ with no real component $\beta > \beta_{\mathrm{c}}(\tau)$, and there are no values of $k_j$ which are purely imaginary if $\beta$ is less than this critical value. Conversely, for fixed $\beta$, there exists a critical value of $\tau$, denoted $\tau_{\mathrm{c}}(\beta)$ such that there are two values of $k_j$ with no real component for $\tau < \tau_{\mathrm{c}}(\beta)$, and no purely imaginary values of $k_j$ if $\tau$ exceeds this critical value. Hence, non-decaying wave behaviour exists in the solution only if $\beta > \beta_{\mathrm{c}}(\tau)$, or equivalentely, $\tau < \tau_{\mathrm{c}}(\beta)$. The wave behaviour in the $\beta$--$\tau$ parameter space is illustrated in Figure \ref{fig:b-tau2}.

As \eqref{eq.singulant1} is a quartic equation, the four solutions may be computed exactly. We denote the solutions according to their asymptotic behaviour in the limit that $\tau \rightarrow 0$ for fixed $\beta$, which corresponds to the gravity wave limit. The four solutions have the behaviour
\begin{align}
k_1 = -\frac{\i}{\beta} + \mathcal{O}(\tau^{1/6}), \qquad k_2 &= -\frac{\i}{\tau^{1/3}} + \frac{\i}{3\beta} + \mathcal{O}(\tau^{1/6}),\qquad k_{3,4} =-\frac{\i \pm\sqrt{3}}{2\tau^{1/3}} + \frac{\i}{3\beta}+ \mathcal{O}(\tau^{1/6}).
\end{align}
For each of these values of $k_j$, we denote the corresponding singulant as $\chi_j$. The exponentially small waves given by the solution $\chi_1$ tend to known gravitational wave behaviour \cite{Chapman3} in this limit, while the amplitude of the remaining waves tends to zero. 

The three wave contributions associated with $\chi_2$, $\chi_3$, and $\chi_4$ correspond to three elastic wave contributions, and are equivalent in the limit that $\beta \to 0$ and $\tau \to 1/\beta$ to those found in \cite{Lustri7}. One of these wave contributions, corresponding to the singulant $\chi_2$, produces waves that do not decay in space away from the obstacle in this limit, while the remaining wave contributions decay exponentially in space. This is consistent with the behaviour identified in \cite{Lustri7}. 

The exact form of these solutions may be determined using a computational algebra package, allowing us to determine the critical values of $\beta$ and $\tau$ exactly. This bifurcation corresponds to a branch point in the exact solutions at $27\beta_{\mathrm{c}}^3 - 256 t_{\mathrm{c}} = 0$, or
\begin{equation}\label{eq.critical}
3\beta_{\mathrm{c}} = 2^{8/3}\tau^{1/3}_{\mathrm{c}}.
\end{equation}
This curve in the $\beta$--$\tau$ parameter space divides solutions with non-decaying oscillations corresponding to gravity waves (corresponding to the solution $k_1$) and elastic waves (corresponding to the solution $k_2$), and solutions in which all four of the wave contributions decay. \textcolor{black}{This result corresponds to setting the flow velocity $U$ to be equal to $c_{\mathrm{min}}$ in \eqref{2.critical}. It is not surprising that we recover the critical speed from the infinite-depth problem, as the linearization step requires the obstacle depth to be small compared to the channel depth. These results are therefore consistent with the phase velocity behaviour \eqref{2.dispersion}.} It is impossible in this regime to choose parameters such that only one of the upstream elastic waves or downstream gravitational waves decay, while the other has constant amplitude. The solution either contains both non-decaying surface wave contributions, or neither. We will see later that this is not necessarily true for nonlinear geometries. 

\textcolor{black}{When non-decaying waves are present, and therefore $k_j$ is imaginary, the surface behaviour can be computed by integrating \eqref{2d:expsm} with respect to $x$. Noting that $\phi = x$ to leading order in $\delta$, $\Lambda_j$ is real when $k_j$ is imaginary, and recalling that the wave amplitude is scaled by $\delta$, we can obtain a wave amplitude of
\begin{equation}\label{2.amplitude}
\mathrm{Amplitude} \sim \frac{2\pi  \Lambda_j \delta}{\beta(1-4\tau \i k_j)}\e^{-|k_j|\pi/\epsilon} \quad \mathrm{as} \quad \eps \to 0.
\end{equation}
From the leading-order behaviour $q^{(0)}$ in \eqref{2.q0lin} it can be determined that the non-dimensional step height is $\delta/2$ to leading order in $\epsilon$. Hence, we are able to determine a relationship between the wave amplitude and the step height in the linearized regime.}

\textcolor{black}{From this expression, the amplitude can be expressed in terms of the upstream flow speed and and the step height, rather than $\delta$ and $\epsilon$. The step height is $\delta/2$ in the limit that $\eps \to 0$. The upstream flow speed can be incorporated using \eqref{2.critical} and \eqref{2.froude} to give $U/c_{\mathrm{min}} = 3^{1/4}/(4\tau\epsilon^3)$, which can be solved for $\eps$. This allows the amplitude expression in \eqref{2.amplitude} to be written in terms of physical properties of the flow geometry, and used to determine quantities such as the wave energy, which scales with the square of the amplitude.}

\subsection{Stokes Structure}

\begin{figure}[tb]
\centering

\subfloat[Analytically-continued free surface: $\beta > \beta_\mathrm{c}$]{
\includegraphics{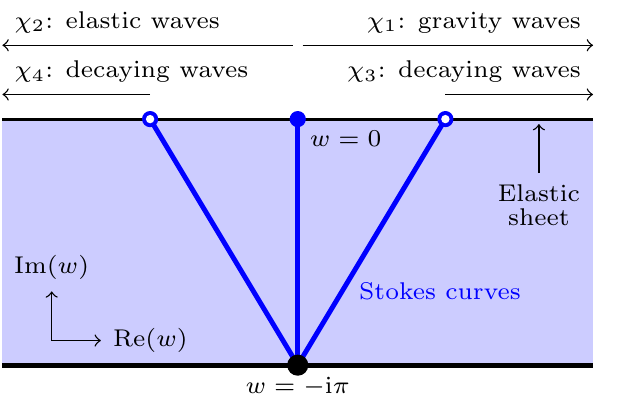}
}
\subfloat[Analytically-continued free surface: $\beta < \beta_\mathrm{c}$]{
\includegraphics{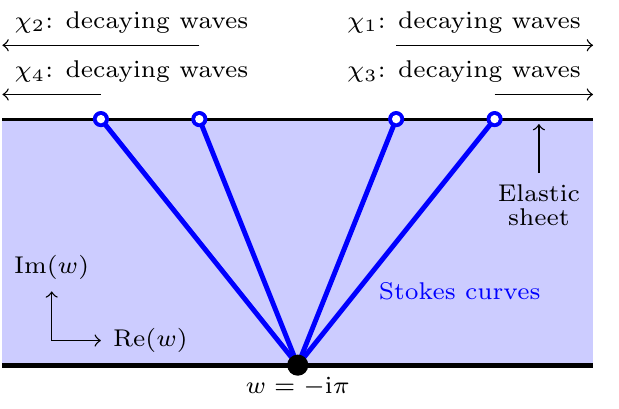}
}

\subfloat[Physical wave profile: $\beta > \beta_\mathrm{c}$]{
\includegraphics{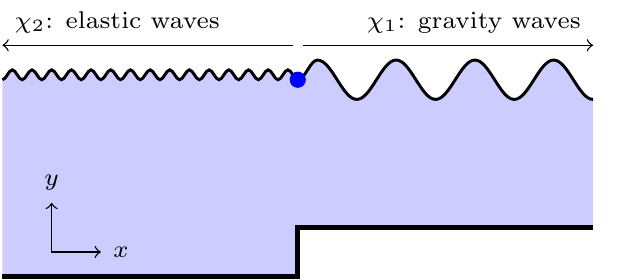}
}
\subfloat[Physical wave profile: $\beta < \beta_\mathrm{c}$]{
\includegraphics{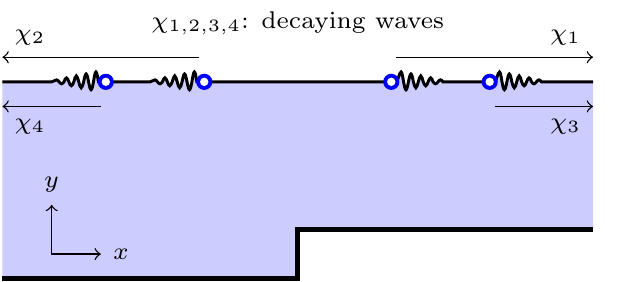}
}

\caption{Schematics of the analytically-continued free surface are presented in (a) and (b). Stokes curves are shown as blue lines that originate at the leading-order singularity, shown as a black circle at $w = -\i\pi$. The points where the Stokes curve intersect the free surface are depicted as blue circles. If the waves are constant amplitude, the blue circles are filled. If the waves decay in space, the circles are unfilled. The regions in which the wave contributions are present are indicated by arrows above the schematic. The cases for $\beta > \beta_{\mathrm{c}}$ and $\beta < \beta{\mathrm{c}}$ are shown in (a) and (b) respectively. In (a), the free surface contains constant amplitude upstream elastic waves and downstream gravity waves, which are both switched on across the Stokes curve that follows $\mathrm{Re}(w) = 0$. The surface intersects two other Stokes curves, which switch on decaying free-surface waves. In (b), the free-surface only contains decaying waves in the far field. This physical behaviour is shown in (c) and (d) for $\beta > \beta_{\mathrm{c}}$ and $\beta < \beta{\mathrm{c}}$ respectively. In (c), the constant-amplitude upstream and downstream waves are depicted. The decaying waves are exponentially small compared to the constant-amplitude waves \textcolor{black}{as $\eps \to 0$}, and therefore not shown. In (d), all four wave contributions are shown, and it can be seen that they decay in space.}\label{FigSL}
\end{figure}

The Stokes structure of the analytically-continued free surface is presented in Figure \ref{FigSL} (a) and (b) for $\beta > \beta_{\mathrm{c}}$ and $\beta < \beta_{\mathrm{c}}$ respectively. Schematics representing the physical flow behaviour are shown in Figure \ref{FigSL} (c) and (d) for $\beta > \beta_{\mathrm{c}}$ and $\beta < \beta_{\mathrm{c}}$ respectively.

Figure \ref{FigSL} (a) shows the Stokes structure on the analytically-continued free surface for  $\beta > \beta_{\mathrm{c}}$. Here, elastic and gravity wave contributions are switched on across a Stokes curve that extends vertically from the singularity at $w = -\i \pi$, corresponding to $\chi_1$ and $\chi_2$ respectively. The gravity waves extend downstream from the Stokes curve, while the elastic waves extend upstream. As $k_1$ and $k_2$ are imaginary, the waves do not decay in space, but rather persist with constant amplitude.

Two other Stokes curves are present on the analytically-continued free-surface, corresponding to $\chi_3$ and $\chi_4$. These contributions cause rapidly decaying waves to appear downstream and upstream from the obstacle respectively. In this case, the direction of propagation is determined by the sign of $\mathrm{Re}(k_3)$ and $\mathrm{Re}(k_4)$. As $\mathrm{Re}(k_3) > 0$, the waves must appear on the downstream side of the Stokes curve, as they would grow exponentially in the upstream direction. Conversely, as $\mathrm{Re}(k_3) < 0$, the waves must appear only in the upstream direction, where they decay exponentially in space.

A schematic of the physical behaviour of this system is shown in Figure \ref{FigSL} (c), which depicts constant-amplitude gravity and elastic waves in the downstream and upstream direction respectively. The decaying waves are exponentially small compared to the constant-amplitude waves \textcolor{black}{as $\eps \to 0$} even at the point in the surface where they first appear, so they are not depicted in the schematic. In this schematic, the gravitational waves are represented with a larger amplitude than the elastic waves. From Figure \ref{fig:b-tau1} (b) and (d), we see that $0 > \mathrm{Im}(k_1) > \mathrm{Im}(k_2)$ in the region $\beta > \beta_{\mathrm{c}}$. From the form of the exponentially small terms in \eqref{2d:expsm}, this implies that the gravity waves associated with $\chi_1$ must have a greater amplitude than the elastic waves associated with $\chi_2$.

Figure \ref{FigSL} (b) shows the Stokes structure on the analytically-continued free surface for  $\beta < \beta_{\mathrm{c}}$. As $\mathrm{Re}(k_j) < 0$ for $j = 1,3$, the waves associated with $\chi_1$ and $\chi_3$ must decay downstream from the corresponding Stokes curve. Conversely, as $\mathrm{Re}(k_j) > 0$ for $j = 2,4$, the associated waves must decay upstream from the corresponding Stokes curve. A schematic of this physical configuration is shown in Figure \ref{FigSL} (d). All four wave contributions decay exponentially in space, meaning that the surface far upstream and downstream from the obstacle must be flat, with no waves present.

\subsection{Nonlinear Problem}\label{S:nonlinear2d}

We note that an exponential asymptotic analysis of gravity-capillary waves in nonlinear regimes revealed a complicated wave structure \cite{Trinh4}, including second-generation Stokes switching \cite{Chapman4}, which was caused by interactions between gravity and capillary effects. We will outline the steps required in order to analyse flexural-gravity waves in a nonlinear regime, and determine the singulant equation for general flow over topography. We will then consider the singulant behaviour for flow over a step in a nonlinear regime. This geometry again corresponds to Figure \ref{Fig1}, although the step height is no longer small.

The formulation of this problem without linearization is largely analogous to the previous analysis, except that the analytically-continued nonlinear dynamic boundary condition is given by \eqref{2.potentialbern}. The remaining steps follow essentially the same format. We obtain a system of recurrence equations for the series terms that is similar to \eqref{eq.1recur1}--\eqref{eq.1recur2}, which can be solved to determine the algebraic series terms for the flow behaviour. The leading order behaviour of the flow is given by
\begin{equation}
\theta^{(0)} = 0,\qquad q^{(0)} = \left(\frac{\e^{-w}+b}{\e^{-w} + 1}\right)^{1/2}.
\end{equation}
We now pose a late-order ansatz identical to \eqref{eq.1ansatz} and apply this to the recurrence relation. Matching the resultant expression in the limit that $n \rightarrow \infty$ gives a singulant equation at leading order,
\begin{equation}\label{eq.nonlinearsing}
1 - \beta \left(q^{(0)}\right)^3 \diff{\chi}{w} + \beta \tau \left(q^{(0)}\right)^4 \left(\diff{\chi}{w}\right)^4 = 0,\qquad \chi(-\i\pi) = 0.
\end{equation}
This is a nonlinear differential equation, which depends on the leading-order flow behaviour. Even without solving this differential equation, we are able to make some observations regarding the upstream and downstream flow behaviour. We denote $q^{(0)}$ in the limit that $w \rightarrow \infty$ as $q_{\mathrm{down}}$, corresponding to the velocity far downstream from the obstacle. Similarly, we denote upstream flow velocity, corresponding $q^{(0)}$ in the limit $w \to -\infty$, as $q_{\mathrm{up}}$. 

\begin{figure}[tb]
\centering

\subfloat[Upwards step: $b = 2$]{
\centering
\includegraphics{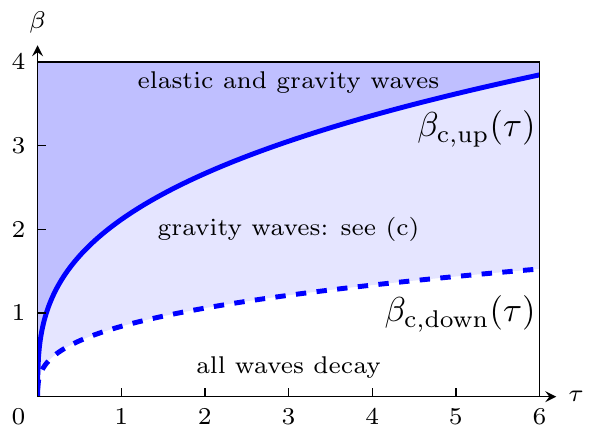}
}
\subfloat[Downwards step: $b = 2$]{
\centering
\includegraphics{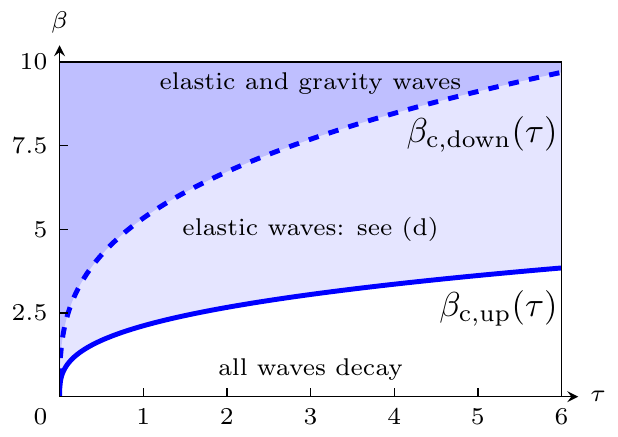}
}

\subfloat[Physical wave profile: $\beta_{\mathrm{c,down}}> \beta > \beta_{\mathrm{c,up}}$]{
\includegraphics{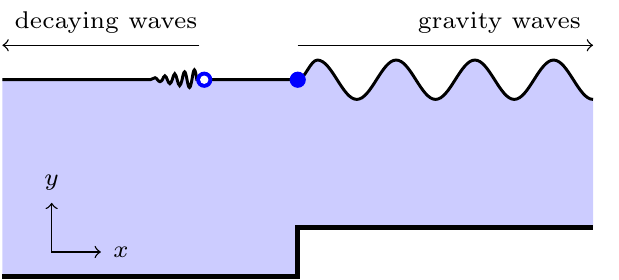}
}
\subfloat[Physical wave profile: $\beta_{\mathrm{c,up}}> \beta > \beta_{\mathrm{c,down}}$]{
\includegraphics{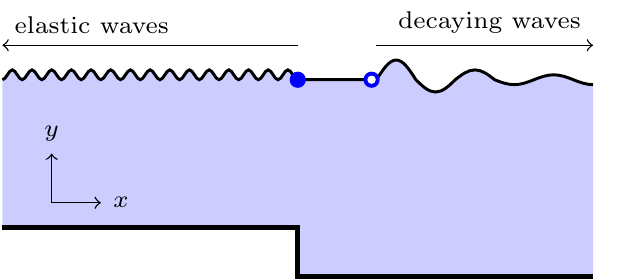}
}

\caption{Figures (a) and (b) depict of the possible wave behaviour as $\beta$ and $\tau$ are varied, for (a) an upwards step with $b = 2$, and (b) a downwards step with $b=2$. In (a) and (b), there are three regions. In the unshaded region, all elastic and gravitational wave contributions must decay in space away from the step. In the dark shaded region, both elastic and gravitational waves can propagate indefinitely with constant amplitude. In (a), the lightly shaded region corresponds to  $\beta_{\mathrm{c, down}} < \beta < \beta_{\mathrm{c, up}}$, and the surface can only support non-decaying waves in the downstream region, driven by gravity. Any upstream elastic waves must decay in space. In (b), the lightly shaded region corresponds to $\beta_{\mathrm{c, up}} < \beta < \beta_{\mathrm{c, down}}$, and the surface can only support non-decaying waves in the upstream region, driven by elastic forces. Figure (c) and (d) depicts the one-sided wave profiles for (a) and (b) respectively. In (c), the schematic shows an upwards step that produces downstream gravity waves with constant amplitude, while the upstream waves decay. In (d), the schematic shows a downwards step that produces upstream elastic waves with constant amplitude, whilethe downstream gravity waves decay. Neither of these two behaviours are possible in the linearized system}\label{fig:b-tau3}
\end{figure}

The nonlinear system has different critical values of $\beta$ and $\tau$ for upstream and downstream waves. The downstream critical values are \textcolor{black}{given} by
\begin{equation}\label{eq.critical_down}
3 q_{\mathrm{down}}^{8/3}  \beta_{\mathrm{c,down}} = 2^{8/3}\tau^{1/3}_{\mathrm{c,down}},
\end{equation}
while the upstream critical values are given by
\begin{equation}\label{eq.critical_up}
3q_{\mathrm{up}}^{8/3} \beta_{\mathrm{c,up}} = 2^{8/3}\tau^{1/3}_{\mathrm{c,up}},
\end{equation}
where the subscripts indicate whether the critical value describes the upstream or downstream region. For the step geometry in Figure \ref{Fig1}, we have $q_{\mathrm{down}} = b^{1/2}$, while $q_{\mathrm{up}} = 1$. This gives
\begin{equation}
\beta_{\mathrm{c,up}} = \left(\frac{256}{27} \tau_{\mathrm{c,up}}\right)^{1/3},\qquad \beta_{\mathrm{c,down}} = \left(\frac{256}{27b^4}\tau_{\mathrm{c,down}}\right)^{1/3}.
\end{equation}
There are three possible elastic sheet behaviours. If $\beta < \beta_{\mathrm{c,down}}$, all waves on the free surface must decay in space away from the step. If $\beta > \beta_{\mathrm{c,up}}$, then the surface can contain non-decaying gravitational waves downstream from the step, and non-decaying elastic waves upstream from the step. These configurations were both possible in the linearized problem. However, if $\beta_{\mathrm{c,down}} < \beta <  \beta_{\mathrm{c,up}}$, then any downstream gravitational waves have non-decaying amplitude, while all elastic effects must decay in space away from the step.  The parameter regimes are illustrated for a step with $b = 2$ in Figure \ref{fig:b-tau3}(a).

It is also possible to consider a downwards step, such that $\theta_0$ = 0 for $-b < \zeta < -1$. In this case, the leading-order solution is given by
\begin{equation}
\theta^{(0)} = 0,\qquad q^{(0)} = \left(\frac{\e^{-w}+1}{\e^{-w} + b}\right)^{1/2}.
\end{equation}
The new critical values are instead given by
\begin{equation}
\beta_{\mathrm{c,up}} =  \left(\frac{256}{27}\tau_{\mathrm{c,up}}\right)^{1/3},\qquad \beta_{\mathrm{c,down}} =\left(\frac{256b^4}{27} \tau_{\mathrm{c,down}}\right)^{1/3}.
\end{equation}
If $\beta < \beta_{\mathrm{c,up}}$, all waves on the free surface must decay in space away from the step. If $\beta > \beta_{\mathrm{c,down}}$, then the flow exceeds both critical values of $\beta$, and can support non-decaying gravitational waves downstream from the step, and non-decaying elastic waves upstream from the step. If $\beta_{\mathrm{c,up}} < \beta <  \beta_{\mathrm{c,down}}$, then any downstream gravitational waves must decay in space, while non-decaying elastic effects are possible. The parameter regimes are illustrated for a downwards step with $b = 2$ in Figure \ref{fig:b-tau3}(b).

\textcolor{black}{This behaviour is consistent with the dispersion relation for finite-depth flow \eqref{2.dispersion_shallow}. We denote the upstream depth of the channel as $L_{\mathrm{u}}$ and the downstream depth as $L_{\mathrm{d}}$; for an upwards step, $L_{\mathrm{d}} < L_{\mathrm{u}}$. This means that the minimum value for the phase velocity, $c_{\mathrm{min}}$ differs on either side of the step. As $L$ increases, it can be seen from the dispersion relation \eqref{2.dispersion_shallow} that the minimum value of the phase speed for flexural--gravity waves also increases. In Figure \ref{fig:b-tau3}(a), we see that the geometry permits an intermediate region in which there are only downstream gravity waves. This corresponds to the case where the flow velocity exceeds $c_{\mathrm{min}}$ in the downstream region, but is lower than the higher value of $c_{\mathrm{min}}$ obtained in the upstream region. The converse is true for downstream steps, where $L_{\mathrm{u}} < L_{\mathrm{d}}$; in this case, the minimum phase velocity for waves is greater in the downstream region than the upstream region, leading to flow geometries described in Figure \ref{fig:b-tau3}(b) which only contain upstream waves.}

This is not a full analysis of the nonlinear problem. We would need to study $\chi$ in order to identify the position of Stokes curves in the problem\footnote{Nonlinear problems can also contain more complicated switching behaviour, such as second-generation Stokes switching \cite{Body1,Chapman4}. These were found to exist in some nonlinear gravity-capillary wave regimes \cite{Trinh4}.}, and therefore determine the location at which the waves appear. This requires determining the solution to \eqref{eq.nonlinearsing}, which would likely necessitate a computational study. Any conclusions reached for the nonlinear system would require validation against numerical simulations, but the presence of surface waves in both directions far from the step makes it challenging to obtain sensible boundary conditions for the flow behaviour. For a detailed description of the numerical challenges involved in studying these systems, and substantial progress in overcoming these obstacles, see the numerical analysis of two dimensional gravity-capillary waves in \cite{Jamshidi1}. A full analysis of the nonlinear problem is therefore beyond the scope of the present study.

\textcolor{black}{An important difference between the analysis of the linear problem and any full nonlinear analysis is that the wavelength of the flexural--gravity waves will depend on $b$, and therefore the step height. This may be seen by the inclusion of $q^{(0)}$ in the singulant equation \eqref{eq.nonlinearsing}. The explicit dependence can be computed by solving \eqref{eq.nonlinearsing} using the asymptotic behaviour of $q^{(0)}$ in the limit that $w \to -\infty$ for upstream waves, and $w \to \infty$ for downstream waves. This phenomenon is predicted by the dispersion relation \eqref{2.dispersion_shallow}, which explicitly depends on the depth of the channel, and is consistent with other related exponential asymptotic studies \cite{Chapman3,Chapman6,Lustri6,Trinh4}. This wavelength selection did not occur in the linearized problem, as the flow was linearized around an unperturbed flow of constant depth; this constant depth determines the wavelength of the flexural--gravity waves to leading order in $\delta$.}

\section{Three-Dimensional Hydroelastic Waves}

\begin{figure}
\centering
\includegraphics{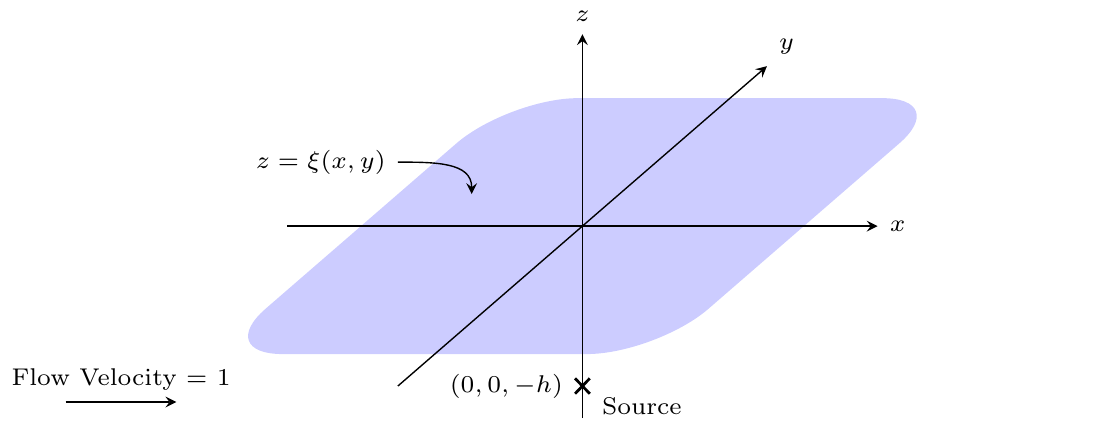}
%
%
%
%

\caption{Prescribed fluid configuration for three-dimensional flow with unit non-dimensionalized velocity past a source with non-dimensionalized depth $h$. The shaded region represents the position of the free surface $\xi(x,y)$, and the cross represents the position of the source. The flow region lies below the free surface, and the mean flow is moving from left to right, with flow velocity $U$ in the unscaled problem. Elastic waves form upstream from the obstacle.
}
\label{1:3dsteadyconfig}
\end{figure}

We consider a three-dimensional incompressible, irrotational, inviscid flow of infinite depth with a submerged point source at depth $H$ and upstream flow velocity $U$. An elastic sheet with flexural rigidity $D$ rests on the surface of the flow.  In three dimensions, we non-dimensionalise velocity by the upstream flow velocity $U$, and distance by a reference length scale $L$. The position of the elastic sheet is denoted as $\xi(x,y)$. The flow therefore has a non-dimensionalized source depth $h = H/L$. 

\begin{figure}
\centering
\includegraphics[width=0.6\textwidth]{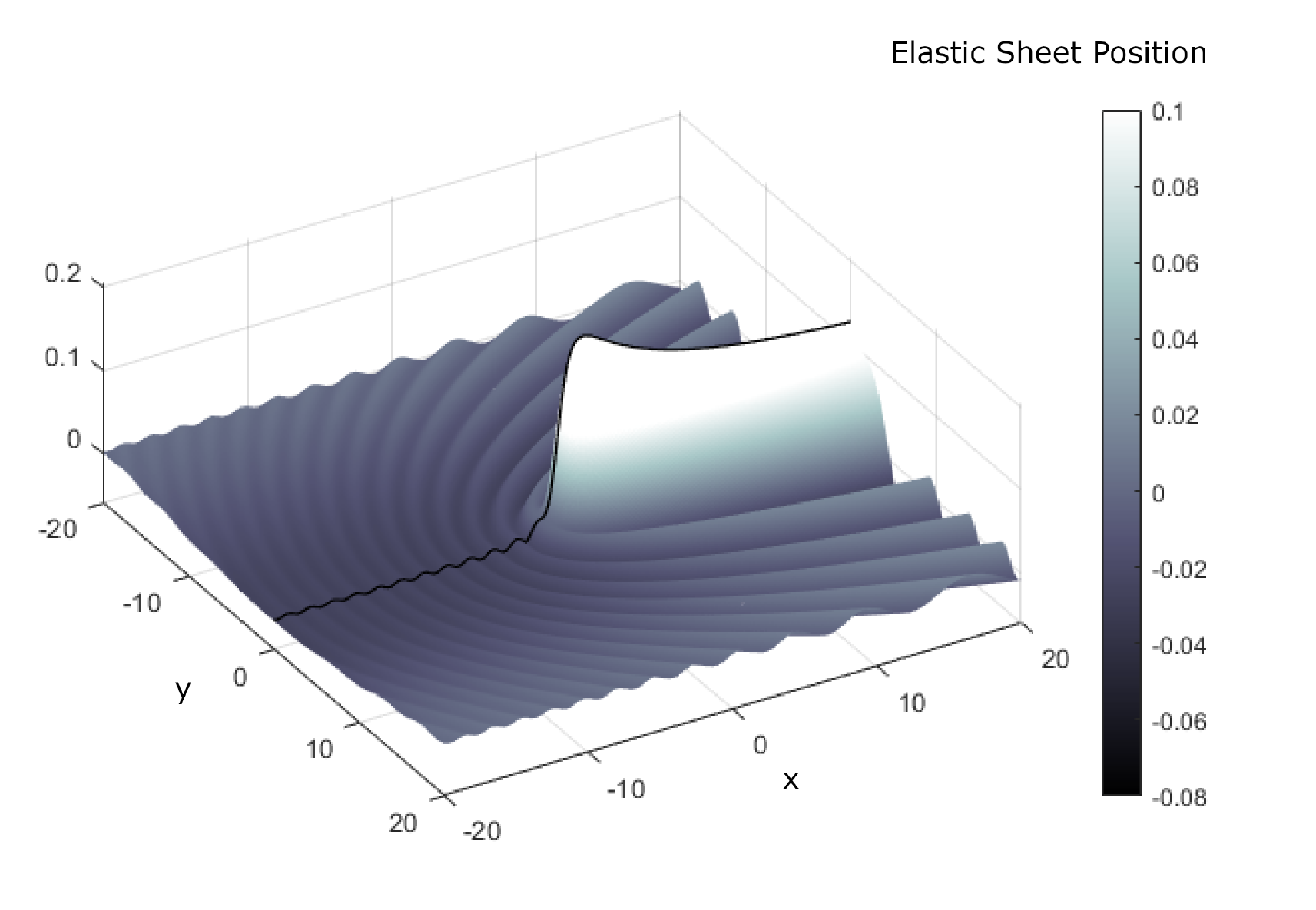}
\caption{Computed \textcolor{black}{three-dimensional linearized flow for $\epsilon = 0.15$ past a source with unit depth, satisfying the system given in \eqref{1:lgeq}--\eqref{1:lsc1}}. The surface along the line $y = 0$ is shown as a black curve, with visible ripples ahead of the source, submerged one unit under $(x,y) = (0,0)$.}
\label{fig:3Dcomp}
\end{figure}

A schematic of this flow geometry is shown in Figure \ref{1:3dsteadyconfig}. A computed elastic sheet solution for $\epsilon = 0.15$, where $\eps^3 = D/(\rho U^2 L^3)$, is shown in Figure \ref{fig:3Dcomp}. The waves that form in the elastic sheet persist upstream from the obstacle, in a similar fashion to capillary waves. This scaling regime neglects gravitational effects; this decision is justified by experimental work such as \cite{Ono1}, in which gravitational effects are present, but the gravitational wavelength is sufficiently large that they are not apparent in the experimental results. Hence, we can predict the behaviour of the hydroelastic waves in such a setup without incorporating gravitational effects.

\subsection{Governing equation}

The flow is governed by Laplace's equation in three dimensions,
\begin{equation}
\label{1:nlgeq} \nabla^2\phi = 0,\qquad -\inf < z < \xi(x,y),
\end{equation}
with kinematic boundary condition  
\begin{alignat}{2}
\label{1:nlbc1}\xi_x\phi_x + \xi_y\phi_y &= \phi_z,\qquad & z &= \xi(x,y).
\end{alignat}
As discussed in Section \ref{S:elastic}, we apply a biharmonic model to describe the behaviour of the elastic sheet in the linear regime. This gives
\begin{alignat}{2}
\label{1:nlbc2}\frac{1}{2}\left(|\nabla\phi|^2-1\right) +\eps^3(\xi_{xxxx} + 2 \xi_{xxyy} + \xi_{yyyy})  &= 0,\qquad & z &= \xi(x,y),
\end{alignat}
where $\epsilon^3 = D/(\rho U^2 L^3)$. \textcolor{black}{This quantity corresponds to the ratio between the Froude number and bending length ratio presented in \eqref{2.froude}. We are concerned with the free-surface behavior in the limit $0 < \eps \ll 1$, corresponding to a regime in which gravity is neglected, and inertial effects are large compared to the elastic restoring force}. Since the flow is uniform in the far field, $\phi_x \rightarrow 1$. The source condition is set to
\begin{equation}
\label{1:nlsc1} \phi \sim \frac{\delta}{4\pi\sqrt{x^2 + y^2 + (z+h)^2}} \qquad \mathrm{as} \quad (x,y,z) \rightarrow (0,0,-h).
\end{equation}
Finally, we can prescribe that the solution satisfies a radiation condition, with waves present directly upstream from the singularity. 

\textcolor{black}{We are concerned with the limit $0 < \delta \ll \eps$, describing a weak source. In this case, the flow disturbance due to the source effect is small, and the equations may be linearized in $\delta$ about a uniform stream while retaining the full asymptotic behaviour in the small $\eps$ limit. We are therefore studying the combined asymptotic parameter regime $0 < \delta \ll \eps \ll 1$.}

\subsection{Linearization}
We linearize about uniform flow by setting
\begin{equation*}
 \phi = x + \delta\tilde{\phi},\qquad \xi = \delta\tilde{\xi},
\end{equation*}
to give, at leading order in $\delta$
\begin{alignat}{2}
\label{1:lgeq} \nabla^2\tilde{\phi} &= 0 ,\qquad &-\inf < z& < 0,\\
\label{1:lbc1}\tilde{\phi}_z - \tilde{\xi}_x &= 0,\qquad & z&= 0,\\
\label{1:lbc2}\tilde{\phi}_x -\eps^3 \left(\tilde{\xi}_{xxxx} +2 \tilde{\xi}_{xxyy} +  \tilde{\xi}_{yyyy}\right)&=0  ,\qquad & z &= 0,
\end{alignat}
where the boundary conditions are now applied on the fixed surface $z = 0$. The far-field conditions imply that $\tilde{\phi} \rightarrow 0$ as $x^2 + y^2 + z^2 \rightarrow \inf$, while near the source, the singular behaviour is given by
\begin{equation}
\label{1:lsc1}\tilde{\phi} \sim \frac{1}{4\pi\sqrt{x^2 + y^2 + (z+h)^2}} \qquad \mathrm{as} \quad (x,y,z) \rightarrow (0,0,-h).
\end{equation}

\subsection{Series expression}\label{SERIES}

We first expand the fluid potential and free-surface position as a power series in $\eps$,
\begin{equation}
 \label{1:series}\tilde{\phi} \sim \sum_{n=0}^{\inf}\eps^{3n}\phi^{(n)},\qquad \tilde{\xi} \sim \sum_{n=0}^{\inf}\eps^{3n}\xi^{(n)},
\end{equation}
to give for $n \geq 0$
\begin{alignat}{2}
\label{1:sergeq} \nabla^2 \phi^{(n)} &= 0,\qquad &-\inf < z& < 0,\\
\label{1:serbc1}\phi^{(n)}_z - \xi^{(n)}_x &= 0,\qquad & z&= 0,\\
\label{1:serbc2}{\phi}_x^{(n)} -{\xi}_{xxxx}^{(n-1)} -2  {\xi}_{xxyy}^{(n-1)} - {\xi}_{yyyy}^{(n-1)} &=0 ,\qquad & z &= 0,
\end{alignat}
with the convention that $\xi^{(-1)} = 0$. The far-field behaviour tends to zero at all orders of $n$, and the singularity condition \eqref{1:lsc1} is applied to the leading-order expression, giving
\begin{equation}
\label{1:sersc1}\phi^{(0)} \sim \frac{1}{4\pi\sqrt{x^2 + y^2 + (z+h)^2}} \qquad \mathrm{as} \quad (x,y,z) \rightarrow (0,0,-h).
\end{equation}
The leading-order solution is given by
\begin{align}
 \label{1:phi0} \phi^{(0)} &= \frac{1}{4\pi\sqrt{x^2 + y^2 + (z+h)^2}} - \frac{1}{4\pi\sqrt{x^2 + y^2 + (z-h)^2}},\\
 \label{1:xi0} \xi^{(0)} &= -\frac{x h}{2 \pi (y^2 + h^2)\sqrt{x^2+y^2+h^2}} - \frac{1}{2\pi(y^2+h^2)},
\end{align}
where the leading-order free surface behaviour is set to be undisturbed far behind the the source.

\subsection{Late-Order Terms}\label{LOT_steady}

In order to optimally truncate the asymptotic series prescribed in \eqref{1:series}, we must determine the form of the late-order terms. To accomplish this, we make a factorial-over-power ansatz with the form
\begin{equation}
 \label{1:ansatz1} \phi^{(n)} \sim \frac{\Phi(x,y,z)\Gamma(3n+\gamma)}{\chi(x,y,z)^{3n+\gamma}}\quad \mathrm{and} \quad \xi^{(n)} \sim \frac{\Xi(x,y)\Gamma(3n+\gamma)}{\chi(x,y,0)^{3n+\gamma}} \quad \mathrm{as} \quad n \rightarrow \inf,
\end{equation}
where $\gamma$ is a constant. In order that \eqref{1:ansatz1} is the power series developed in Section \ref{SERIES}, we require that the singulant, $\chi$, satisfies
\begin{equation}
 \label{1:singularitycondition}  \chi = 0 \qquad \mathrm{on} \qquad x^2 + y^2 + (z\pm h)^2 = 0,
\end{equation}
where the sign chosen depends upon which of the two singularities is being considered. For complex values of  $x$, $y$ and $z$, this defines a four-dimensional hypersurface. Irrespective of which singularity is under consideration, this hypersurface intersects the four-dimensional complexified free surface on the two-dimensional hypersurface satisfying $x^2 + y^2 + h^2 = 0$.  

\subsubsection{Calculating the singulant}\label{CH5_SINGULANT1}

Applying the ansatz expressions in \eqref{1:ansatz1} to the governing equation \eqref{1:sergeq} and taking the first two orders as $n \rightarrow \inf$ gives, for $z \leq 0$,
\begin{align}
\label{1:ansgov1} \chi_x^2 + \chi_y^2 + \chi_z^2 &= 0,\\
\label{1:ansgov2} 2\Phi_x\chi_x + 2\Phi_y\chi_y + 2\Phi_z\chi_z &= -(\chi_{xx}+\chi_{yy}+\chi_{zz}),
\end{align}
while the boundary conditions on $z = 0$ at leading order become
\begin{align}
 \label{1:ansbc1} -\chi_z\Phi + \chi_x \Xi &= 0,\\
\label{1:ansbc2} \chi_x\Phi + (\chi_x^4 +2 \chi_x^2\chi_y^2  + \chi_y^4)\Xi  &= 0.
\end{align}
The system in \eqref{1:ansbc1}--\eqref{1:ansbc2} must have nonzero solutions, which requires
\begin{equation}\label{1:phixi}
\chi_z = -\frac{\chi_x^2}{(\chi_x^2+\chi_y^2)^2}, \qquad \Xi = -\frac{\chi_x}{\chi_x^2+\chi_y^2} \Phi.
\end{equation}
Applying \eqref{1:phixi} to \eqref{1:ansgov1} evaluated on $z=0$ gives a singulant equation for $\chi$ on the free surface,
\begin{equation}
\label{1:eik}  \chi_x^4 + \left(\chi_x^2 + \chi_y^2\right)^5 = 0.
\end{equation} 
This expression is similar to the capillary wave singulant equation from \cite{Lustri7}, with a different power in the second term. The subsequent analysis is therefore similar, and we include an outline of the details. Because the singularity lies below the fluid surface, we must solve \eqref{1:eik} for complex $x$ and $y$ with the boundary condition
\begin{equation}
\label{1:eik2}  \chi = 0 \quad \mathrm{on} \quad x^2 + y^2 + h^2 = 0.
\end{equation}
Solving \eqref{1:eik}--\eqref{1:eik2} using Charpit's method gives
\begin{equation}
 \label{1:sschiray2} \chi = \pm\frac{3 z_j h^{1/3}s^{5/3}(s-x)}{2h^2+5s^2},
\end{equation}
where $z_j$ is one of the three solutions to $z_j^3 = 1$, and $s$ one of the four solutions to
\begin{equation}
 \label{1:sspoly} 25(x^2+y^2)s^4 + 20h^2 x s^3 + (4h^2+25x^2+20y^2)h^2 s^2 + 20h^4 x s + 4 h^4 (h^2+y^2) = 0.
\end{equation} 
This produces twenty-four potential late-order contributions, corresponding to the choice of sign and $z_j$ in \eqref{1:sschiray2} and the four solutions to \eqref{1:sspoly}. Twelve of these solutions are spurious, introduced by squaring both sides of an equation in the algebraic manipulations. This leaves twelve solutions, which appear as six complex conjugate pairs. Three of these pairs do not demonstrate Stokes switching, as they are exponentially large on the curve $\mathrm{Im}(\chi) = 0$; they must therefore be inactive on the surface, and do not contribute to the free-surface behaviour. This leaves three singulant pairs that produce waves on the free surface. We will denote these as $\chi_j$ and $\overline{\chi}_j$ for $j = 1, 2$, and $3$, where the bar represents complex conjugation.

\begin{figure}
\centering
\subfloat[$\mathrm{Re}(\chi_1)$]{
\centering
\includegraphics{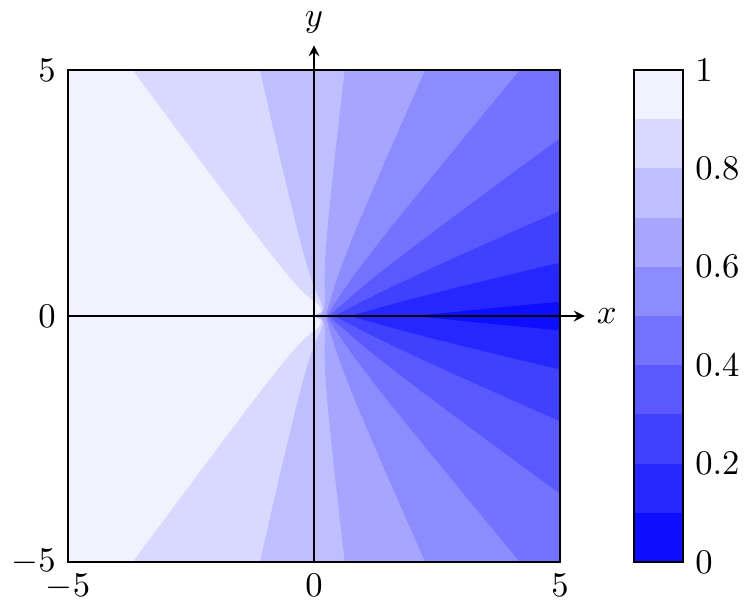}
}
\subfloat[$\mathrm{Im}(\chi_1)$]{
\centering
\includegraphics{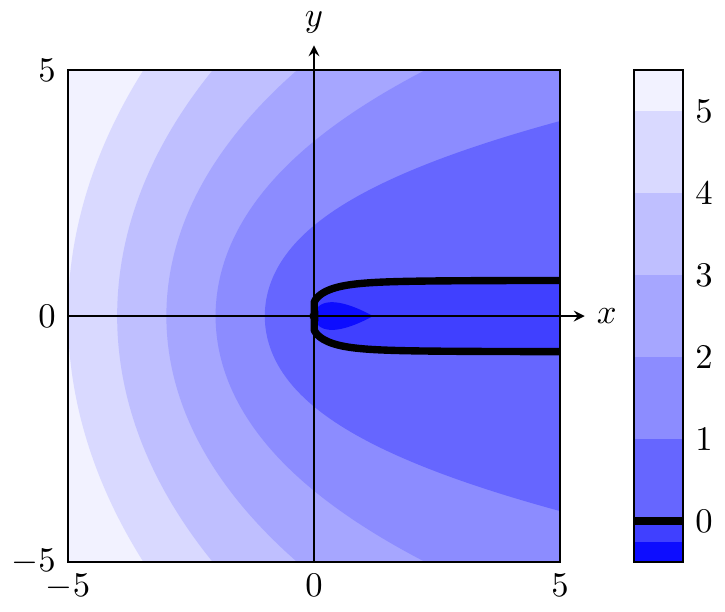}
}
\caption{The (a) real and (b) imaginary part of the singulant $\chi_1$, corresponding to algebraically-decaying elastic waves. The Stokes curve, satisfying $\mathrm{Re_1}(\chi) > 0$ and $\mathrm{Im}(\chi_1) = 0$, is depicted in the second figure. The waves are not present in a region downstream from the obstacle, and appear as the Stokes curve is crossed into the region ahead of the obstacle. The equal phase lines in (b) illustrate the shape of the surface waves.}
\label{f:3D_Stokes}
\end{figure}

The behaviour of $\chi_1$ is illustrated in Figure \ref{f:3D_Stokes}. The surface waves are absent directly downstream from the obstacle. The surface contains a Stokes curve that passes through the origin. This Stokes curve causes elastic waves to be switched on upstream from the obstacle. From direct algebraic computation, we find that $\chi \sim (1-\i x)/h$ in the limit that $x \rightarrow -\infty$. These elastic waves do not decay exponentially in space, although they will have algebraic spatial decay due to the prefactor, calculated below.

\begin{figure}
\centering
\subfloat[$\mathrm{Re}(\chi_2)$]{
\centering
\includegraphics{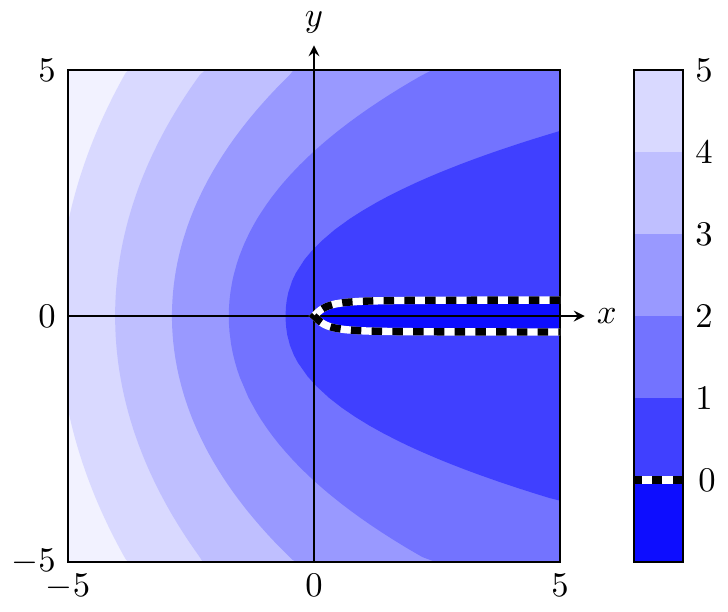}
}
\subfloat[$\mathrm{Im}(\chi_2)$]{
\centering
\includegraphics{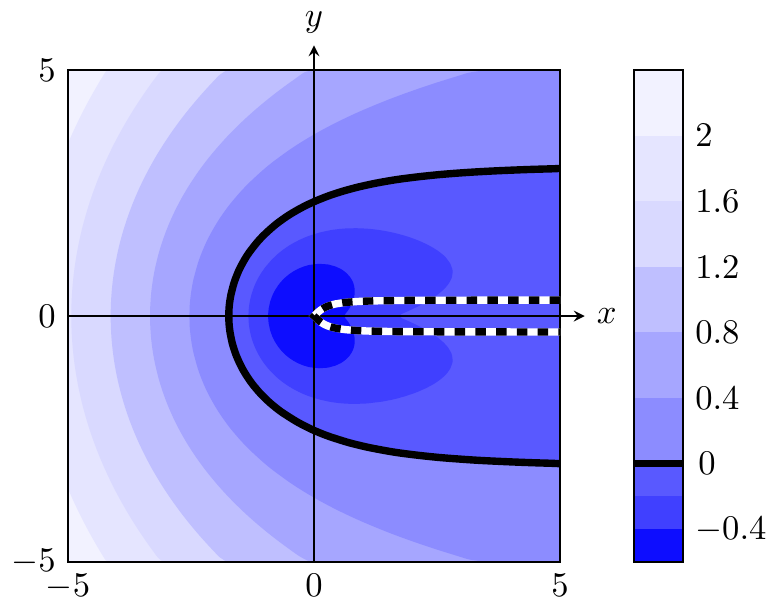}
}
\caption{The (a) real and (b) imaginary part of the singulant $\chi_2$, corresponding to one set of exponentially-decaying elastic waves. The anti-Stokes curve satisfies $\mathrm{Re}(\chi_2) = 0$, and are depicted in both figures as a dashed line. On the inside of this curve (where $\mathrm{Re}(\chi_2) < 0$), the exponential term would be large. Consequently, the remainder must be inactive in a region containing the anti-Stokes curve. The Stokes curve, satisfying $\mathrm{Re}(\chi_2) > 0$ and $\mathrm{Im}(\chi_2) = 0$, is depicted in the second figure. We see that the exponential must be switched on as this curve is crossed in a direction heading away from the origin. As the real part of $\chi_2$ increases without bound as $x \rightarrow -\infty$, these waves must decay exponentially in space, which is too rapid to have an observable physical effect. The behaviour of $\chi_3$ can be obtained by the mapping $ \mapsto -x$.}
\label{f:3D_Stokes2}
\end{figure}

The behaviour of $\chi_2$ is illustrated in Figure \ref{f:3D_Stokes2}. The surface waves are also absent directly downstream from the obstacle. In fact, we see that this does not depend on the radiation condition; instead, there is an anti-Stokes curve on the surface. If the surface waves were present on this side of the Stokes curve, they would become exponentially large on the downstream side of the anti-Stokes curve. We therefore see that the waves are present only on the downstream side of the Stokes curve. Importantly, $\mathrm{Re}(\chi)$ grows monotonically in the negative $x$ direction, becoming arbitrarily large as $x \rightarrow -\infty$. This means that the waves decay exponentially in space. We will therefore not consider these waves in subsequent analysis.

Finally, we find that $\chi_3$ is identical to $\chi_2$ reflected around the $y$-axis. This is consistent with the two-dimensional result, seen in \cite{Lustri7}, in which the surface contains symmetric waves emerging in both directions above the obstacle, which decay exponentially in space.

\subsubsection{Calculating the prefactor}\label{S:prefactor}

We follow a similar analysis to previous work on gravity and capillary waves in order to determine the prefactor expression $\Phi$, and hence $\Xi$. These calculations are quite technical, and the details are included in Appendix \ref{PREFACTOR_APP}. The prefactor $\Phi$ is given by 
\begin{equation}
 \label{1:Phiss} \Phi= \frac{s^{1/3}\sqrt{2}}{4\pi^{3/2} h^{5/6}}\left[1-\frac{5 h^2 (2 h^2 - s^2) (s - x)}{3s (2 h^2 + 5 s^2) }\right]^{\frac{(-1)^{1/3} (4h^{10}-5h^8s^2-30h^2s^8-48s^{10})}{30 h^{8/3}s^{16/3}(2h^2-s^2)}},\end{equation}
where $s$ is the solution of \eqref{1:sspoly} corresponding to the singulant illustrated in Figure \ref{f:3D_Stokes}.

Finally, to find $\gamma$, we ensure that the strength of the singularity in the late-order behaviour $\phi^{(n)}$, given in \eqref{1:ansatz1} is consistent with the leading-order behaviour $\phi^{(0)}$, which has strength $1/2$. It is clear from the recurrence relation \eqref{1:serbc2} that the strength of the singularity will increase by three between $\phi^{(n-1)}$ and $\phi^{(n)}$. This implies that near the singularity at $x^2 + y^2 + h^2 = 0$,
\begin{equation}
\label{1:gammamatch}\frac{\Phi\Gamma(\gamma)}{\chi^{\gamma}} \rightarrow \frac{\alpha(x,y)}{(x^2+y^2+h^2)^{1/2}},
\end{equation}
where $\alpha$ is of order one in the limit. From \eqref{1:Phiss}, we see that the prefactor is also order one in this limit. A local analysis near the singularity (performed in \ref{1:innerchi1}) shows that  $1/\chi$ contains a singularity with strength one at $x^2 + y^2 + h^2 = 0$. Matching the order of the expressions in \eqref{1:gammamatch} therefore gives $\gamma = 1/2$. We have therefore completely described the late-order terms in \eqref{1:ansatz1}, where \eqref{1:phixi} is used to determine the value of $\Xi$, and hence the behaviour of the free-surface waves. 

In Appendix \ref{CH5_STEADYSMOOTHING}, we use the late-order terms in \eqref{1:ansatz1} to apply the matched asymptotic expansion methodology of \cite{Daalhuis1}. We optimally truncate the asymptotic series and identify the Stokes curves. Finally, we use a matched asymptotic expansion analysis on the truncation remainder to compute the exponentially small contribution to the free surface behaviour that appears across the Stokes lines. This analysis in Appendix \ref{CH5_STEADYSMOOTHING} follows similar steps to the equivalent analysis in \cite{Lustri2,Lustri7}.

Using this method, we find that the exponentially small contributions to the fluid potential (denoted $\phi_{\mathrm{exp}}$) and free surface position (denoted $\xi_{\mathrm{exp}}$) \textcolor{black}{as $\eps \to 0$} are switched in the region to the left of the Stokes curve shown in Figure \ref{f:3D_Stokes}. In the region where the exponentially small contributions are present, they are given by
\begin{equation}
\phi_{\mathrm{exp}} \sim \frac{2\pi\i \Phi}{3\sqrt{\epsilon}}\e^{-\chi_1/\epsilon} + \mathrm{c.c.},\qquad \xi_{\mathrm{exp}} \sim \frac{2\pi\i \Xi}{3\sqrt{\epsilon}}\e^{-\chi_1/\epsilon} + \mathrm{c.c.},\label{1:RnSn}
\end{equation}
where c.c. denotes the complex conjugate contribution. In particular, the expression for $\xi_{\mathrm{exp}}$ contains exponentially small oscillations \textcolor{black}{as $\eps \to 0$} representing the elastic ripples on the free surface. 

We note that is a leading-order expression for the exponentially small waves, with algebraic corrections to the prefactors $\Phi$ and $\Xi$ omitted. The solution also contains contributions from $\chi_2$ and $\chi_3$ with the same form, but these decay exponentially in space, and are therefore exponentially smaller in amplitude \textcolor{black}{as $\epsilon \to 0$} than the elastic waves caused by $\chi_1$, as well as the neglected correction terms. We therefore omit these contributions from the asymptotic expression.

\subsection{Results and Comparison}

Along the curve $y=0$ for $x < 0$, we have $s = \i h$ and $\chi =h + \i x$. We therefore evaluate the free surface position to be
\begin{equation}
\xi_{\mathrm{exp}} \sim -\frac{ (-1)^{1/6} 3^{3/10}}{3  h^{1/5} \sqrt{2\pi\epsilon}  (8h + 5 \i x)^{3/10}}\e^{-(h + \i x)/\epsilon} + \textrm{c.c.}  \quad \mathrm{as} \quad \epsilon \to -\infty,
\end{equation}
where c.c. denotes the complex conjugate contribution. In the limit that $x$ becomes large and negative, we find that the amplitude of the waves on $y = 0$ is is given by
\begin{equation}
\mathrm{Amplitude} \sim \frac{1}{3 h^{1/5}}\sqrt{\frac{2}{\pi\epsilon}} \left(\frac{3}{5|x|}\right)^{3/10}\e^{-h/\epsilon}	 \quad \mathrm{as} \quad x \to -\infty,\, \epsilon \rightarrow 0.\label{eq.amp}
\end{equation}
This provides us with a quantity we may use to check the accuracy of the asymptotic approximation. We compare the amplitude of the asymptotic results with those of numerically-calculated free surface profiles \textcolor{black}{obtained by solving the linearized system \eqref{1:lgeq}--\eqref{1:lsc1}. These results were obtained using an adaptation of the method described \cite{Lustri2, Lustri7}, which consists of expressing the free-surface behaviour in terms of Fourier inversion integrals, and evaluating the double integral numerically on a fixed domain}.

\begin{figure}
\centering
\includegraphics{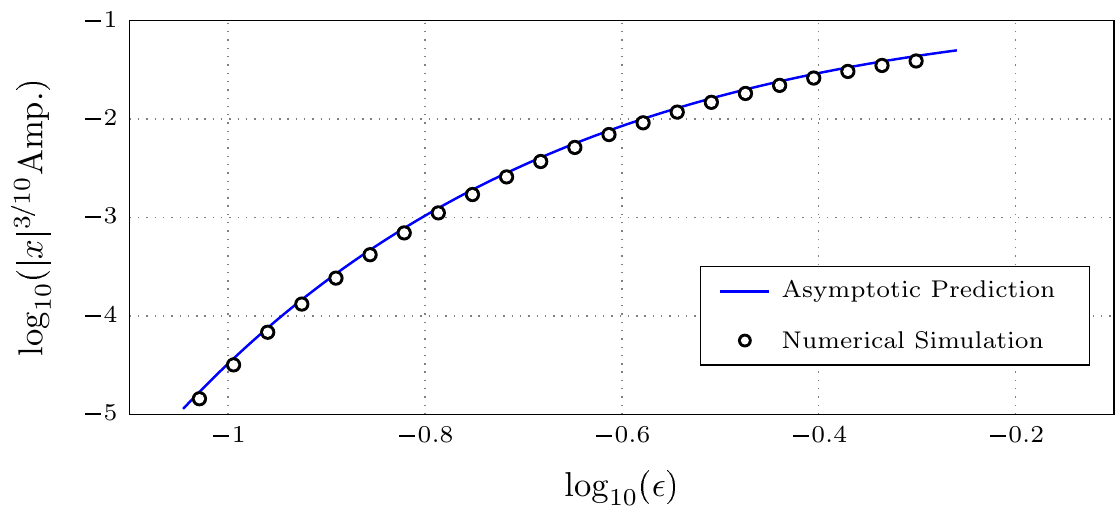}
%
%
%
%
%
%

\caption{Comparison between asymptotic predictions of the amplitude using \eqref{eq.amp} and numerical calculations, for the case $h=1$. The amplitude is scaled by $|x|^{3/10}$ so that it tends to a constant value as $x \rightarrow -\infty$. }\label{fig:3Dnum}
\end{figure}

In Figure \ref{fig:3Dnum}, we illustrate the scaled numerical amplitude (circles) against the asymptotic prediction from \eqref{eq.amp}, computed for $h = 1$ over a range of $\epsilon$ values. The amplitude is scaled by $|x|^{3/10}$, so that it tends to a constant as $x \to -\infty$. The numerical amplitude is taken by determining the scaled amplitude for sufficiently large negative values of $x$ that the scaled amplitude does not display significant variation. There is strong agreement between the asymptotic predictions and numerical results. For values of $\epsilon$ smaller than those depicted, it become numerically challenging to compute the wave behaviour, due to the very small amplitude of the resulting waves.

\section{Discussion and Conclusions}

In this study, studied behaviour of the waves that form on an elastic sheet resting on an invisicid flow stream containing a submerged obstacle in several different regimes. In each regime we considered, the surface waves are exponentially small in the limit of small bending stiffness, and we therefore required exponential asymptotic techniques in order to study the wave behaviour. As these waves share many similarities with surface-tension driven capillary waves, we used techniques applied in \cite{Trinh3,Trinh4} for two-dimensional gravity-capillary waves, and \cite{Lustri7} for three-dimensional capillary waves. Using exponential asymptotics, identified the regions in which the surface waves formed, and calculated a mathematical expression for the wave behaviour. In each flow configuration we determined that the surface waves appeared as Stokes curves were crossed, and that by understanding the behaviour of these Stokes curves, it was possible to classify the types of waves that could appear on the elastic sheet.

We first studied the behaviour of flexural-gravity waves on linearized flow over a small step in two dimensions. In this regime, the elastic sheet behaviour depended on a particular paremeter $\beta_c(\tau)$, which related the relative size of the Froude number and bending stiffness parameter. If the parameter $\beta$ is less than this critical value, the four different wave contributions on the elastic sheet decay away from the obstacle, meaning that both the upstream and downstream regions do not contain any waves. If $\beta$ exceeds this critical value, constant-amplitude elastic waves propagate upstream from the obstacle, and constant-amplitude gravity waves propagate downstream. In this regime the sheet behaviour also contains two decaying wave contributions, but these are small compared to the constant-amplitude waves.

We did not perform a full exponential asymptotic analysis of the two-dimensional nonlinear problem, but instead used the singulant behaviour and Stokes structure of the problem to predict the types of wave that could form on the elastic sheet. Flow over a submerged step in this case contains a third intermediate regime, in addition to the two regimes from the linearized problem. For flow over an upwards step, there are two distinct critical values; one critical value of $\beta$ which determines whether constant-amplitude downstream gravity waves are preesnt, and a larger critical value of $\beta$ which determines whether constant-amplitude upstream elastic waves are present. It is therefore possible to construct flow geometries which contained constant-amplitude downstream gravity waves, but that all upstream waves decay in space. The converse is true for a downwards step; it is possible to construct geometries which contain constant-amplitude upstream elastic waves, but all downstream waves decay in space. 

In the two-dimensional geometries considered in this study, we considered the behaviour of systems in which gravitational and elastic effects have similar strength, while neglecting surface tension effects. It would also be interesting to calculate the wave behaviour while incorporating interactions between \textcolor{black}{surface tension of the fluid} and bending effects. This interaction was studied experimentally in \cite{deike2013nonlinear,deike2017experimental}, and a similar form has been used to study compressive effects on elastic sheets \cite{das2018dynamics}. In this case, \textcolor{black}{both capillary and elastic waves} would be expected to propagate upstream, and the behaviour in the elastic sheet that would be more straightforward to validate computationally.

Finally, we studied hydroelastic waves that form on linearized flow in three dimensions over a submerged obstacle. We considered a regime in which the wave behaviour is caused solely by the elastic restoring force. Using exponential asymptotics, we calculated the wave behaviour, and found that is consistent with the two-dimensional case from \cite{Lustri6}. The flow contained four wave contributions, two of which decay exponentially in space, and one of which produced visible upstream elastic waves that decay algebraically. We validated these calculations against numerical results. The wave patterns are visible similar to the capillary waves from \cite{Lustri7}, although the algebraic decay rate of the wave amplitude differs between elastic and capillary waves, and show qualitative agreement with the experimental results of \cite{Ono1}.

It is natural to consider whether this analysis can be extended to flexural-gravity waves in three dimensions. In this case, we would replace \eqref{1:nlbc2} with
\begin{alignat}{2}
\label{1:nlbcC}\frac{\beta\epsilon}{2}\left(|\nabla\phi|^2-1\right) +\xi + \beta\tau\eps^4(\xi_{xxxx} + 2 \xi_{xxyy} + \xi_{yyyy})  &= 0,\qquad & z &= \xi(x,y),
\end{alignat}
where $F^2 = \beta \eps$ and $\sigma = \beta \tau \eps^4$, as in the two-dimensional problem. For the linearized problem, repeating the late-order analysis in a similar fashion gives
\begin{equation}
\beta \chi_x^4 + (\chi_x^2 + \chi_y)^2[1 + \beta\tau(\chi_x^2 + \chi_y^2)^2]^2 = 0,
\end{equation}
with the boundary condition $\chi = 0$ on $x^2 + y^2 + h^2 = 0$. This equation is challenging to solve using analytical methods, as there are eight valid solution sheets in the analytically-continued free surface, which has $x\in\mathbb{C}$ and $y\in\mathbb{C}$. Solving this equation directly likely require more complicated ray-tracing methods such as those developed for nonlinear three-dimensional flow in \cite{Johnson1}, and would therefore be unlikely to identify convenient closed-form solutions such as \eqref{1:RnSn}. Nonetheless, these computations would likely still be useful, as exponential asymptotic methods are a convenient method for isolating particular wave behaviours and studying the waves directly.

\section{Declaration of Interests}

The author reports no conflict of interest.

\section{Acknowledgements}

CJL acknowledges ARC Discovery Project DP190101190. CJL would like to thank Prof. Scott McCue and Prof. Michael Meylan for useful discussion.

\appendix

\section{Detailed Analysis for Two-Dimensional Geometry}

\subsection{Inner Expansion Near Singularity}\label{S:2Dinner}

Define a new variable $\mu$ such that $w + \i \pi = \epsilon \mu$. To leading order as $\eps\to 0$, \eqref{2.zetacauchy} becomes 
\begin{equation}\label{2.innertheta}
{q} - \i {\theta} \sim \frac{1}{2\eps \mu}.
\end{equation}
Noting the form of $q_0$, we set $\overline{q}(\mu)/2\eps = q(\zeta)$, $\overline{\theta}(\mu)/2\eps = \theta(\zeta)$. We express the Bernoulli condition in terms of $\mu$, and use \eqref{2.innertheta} to eliminate $\theta$ and obtain an equation for $q$. To leading order as $\eps \rightarrow 0$, this gives
\begin{equation}
-\i\beta \overline{q}' + \frac{1}{\mu} - \overline{q} + \beta\tau\left(\frac{24}{\mu^5} - \overline{q}''''\right) = 0.
\end{equation}
Now we create the inner expansion
\begin{equation}
\overline{q} \sim \sum_{n=0}^{\infty} \frac{A_n \Gamma(n+1)}{\mu^{n+1}}\quad \mathrm{as} \quad \mu \to \infty.\label{a:innerexp}
\end{equation}
Applying this to the inner equation and matching powers of $\mu$ gives the following recurrence relation
\begin{align}
1 - A_0 &= 0,\\
\i \beta A_{n-1} - A_n &= 0, \qquad n = 1, 2, 3,\\
\i \beta A_3 - A_4 + \beta \tau (1 - A_0)&= 0,\\
\i\beta A_{n-1} - A_n - \beta\tau A_{n-4} &= 0,\qquad n > 4.\label{2.inner_gen_recur}
\end{align}
Calculating the first few terms gives
\begin{gather}
A_0 = 1,\quad A_1 = \i\beta,\quad A_2 = -\beta^2,\quad A_3 = -\i\beta^3,\quad A_4 = \beta^4,\quad A_5 = \i\beta^2(\beta^3 - \tau), \\A_6 = -\beta^3(\beta^3-2\tau),\quad
A_7 = -\i \beta^4 (\beta^3- 3  \tau),\quad A_8 = \beta^5(\beta^3 - 4\tau).
\end{gather}
While the terms increase in complexity beyond this point, it is possible to solve \eqref{2.inner_gen_recur} exactly, giving
\begin{equation}
A_n = \frac{C_1}{k_1^n} +\frac{C_2}{k_2^n} +\frac{C_3}{k_3^n} +\frac{C_4}{k_4^n},
\end{equation}
where $C_j$ are constants to be determined. We choose four values $A_j$ with $j > 4$, such as $j = 5,6,7,8$, and use these values to find $C_j$ for $j = 1,2,3,4.$ The resultant expressions are given by
\begin{equation}
C_j = \frac{k_j^8(k_j A_5 + k_j^2 A_6 + k_j^3 A_7 -\tfrac{1}{\beta\tau} A_8 )}{ \tfrac{1}{\beta\tau} - 3 k_j^4},
\end{equation}
for $j = 1,\ldots,4$. Using Van Dyke's matching principle to maintain consistency between the behaviour of the inner expansion \eqref{a:innerexp} in the limit $\mu \to \infty$ with the late-order term ansatz \eqref{eq.1ansatz} in the limit that $w \to -\i\pi$, it may be seen that $\Lambda_j = C_j/2$ for $j = 1,2,3,4$, where $\Lambda_j$ is the prefactor corresponding to $\chi_j$. This is not a particularly useful result, although we do note that $C_3 = C_4$. It can also be seen by direct substitution that $C_1$ tends to the correct gravity wave prefactor in the limit that $\tau \rightarrow 0$.
\subsection{Exponential Asymptotic Analysis}\label{S:2Dsmoothing}

We truncate the divergent series after $N-1$ terms, obtaining
\begin{equation}
\hat{q}(w) = \sum_{n=0}^{N-1}\eps^n q^{(n)}(w) + R_N(w),\qquad \hat{\theta}(w) = \sum_{n=0}^{N-1}\eps^n \theta^{(n)}(w)  + S_N(w),
\end{equation}
where $R_N(w)$ and $S_N(w)$ are the remainder terms after optimal truncation. Typically, the optimal truncation point can be found using a heuristic
from \cite{Boyd1}, in which the series is truncated after the smallest term. Approximating series terms using the late-order ansatz gives $N \sim |\chi|/\eps$ as $\eps \rightarrow 0$ and $N \rightarrow \infty$. We therefore set $N = |\chi|/\eps + \alpha$, where $0 \leq \alpha < 1$ is chosen so that $N$ is an integer.

Applying the truncated series to the integral equation and neglecting the integral expression (see Trinh) gives $R_N = \i S_N$. Applying the truncated series to the Bernoulli condition gives
\begin{equation}\label{2.SNeqn}
\i \beta \eps \diff{S_N}{\zeta} + S_N + \beta\tau\eps^4 \diff{^4 S_N}{\zeta^4} \sim \eps^N \theta_N,
\end{equation} 
where the recurrence relation was used to simplify the right-hand side of this expression. The right-hand side is exponentially small except in the neighbourhood of Stokes lines. If we use a WKB ansatz on this problem, neglecting the right-hand side entirely, we determine that the remainder behaviour away from the Stokes line is given by $S_N \sim A \Theta(\zeta)  \e^{-\chi(\zeta)/\eps}$ as $\eps \to 0$, where $A$ is some constant. In order to capture the effect of Stokes switching in the neighbourhood of a Stokes line, we set
\begin{equation}
S_N \sim A(\zeta) \Theta(\zeta) \e^{-\chi(\zeta)/\eps} \quad \mathrm{as} \quad \eps \to 0,
\end{equation}
where $A$ is a Stokes multiplier that varies rapidly in the neighbourhood of a Stokes line, but is essentially constant away from this neighbourhood. Applying this to \eqref{2.SNeqn} and simplifying the resultant expression gives
\begin{equation}
\beta\left(\i + 4 \tau k_j^3\right)\diff{A}{w} \sim \eps^{N-1}\frac{\Gamma(N+1)}{\chi^{N+1}} \e^{-\chi/\eps} \quad \mathrm{as} \quad \eps\to 0.
\end{equation}
We write the solution as a function of the independent variable $\chi$, which gives
\begin{equation}
\frac{\beta}{k_j}\left({\i + 4 \tau k_j^3}\right)\diff{A}{\chi} \sim \eps^{N-1}\frac{\Gamma(N+1)}{\chi^{N+1}} \e^{-\chi/\eps} \quad \mathrm{as} \quad \eps\to 0.
\end{equation}
Now we make the transformation $\chi = r \e^{\i\theta}$, and consider the variation in the $\theta$ direction. Hence, we have
\begin{equation}
\diff{}{\chi} = -\frac{\i\e^{-\i\theta}}{r}\diff{}{\theta}.
\end{equation}
Using Stirling's formula and the optimal truncation $N = r/\eps + \alpha$ on the resultant expression gives
\begin{equation}
\frac{\beta}{k_j}\left({\i + 4 \tau k_j^3}\right)\diff{A}{\theta} \sim \frac{\i\sqrt{2\pi r}}{\eps^{3/2}}\mathrm{exp}\left(\frac{r}{\eps}(\e^{\i\theta} - 1) - \i\theta\left(\frac{r}{\eps} + \alpha\right)\right) \quad \mathrm{as} \quad \eps \to 0.
\end{equation}
To investigate the rapid variation in the neighbourhood of this expression in the neighbourhood of the Stokes line, we set $\theta = \eps^{1/2}\vartheta$, which gives
\begin{equation}
\frac{\beta}{k_j}\left({\i + 4 \tau k_j^3}\right)\diff{A}{\vartheta}\sim \frac{\i\sqrt{2\pi r}}{\eps}\e^{-r\vartheta^2}{2} \quad \mathrm{as} \quad \eps \to 0,
\end{equation}
so that
\begin{equation}
A \sim \frac{k_j}{\beta}\left(\frac{1}{1 - 4 \tau \i k_j^3}\right)\frac{\sqrt{2\pi r}}{\eps}\int_{-\infty}^{\theta\sqrt{r/\eps}} \e^{-t^2/2}\d t + C,
\end{equation}
where $C$ is a constant. As we move from the waveless region across a Stokes line, the jump in the Stokes switching term is given by
\begin{equation}
\left[\mathcal{S}\right]_-^+ \sim\left(\frac{1}{1 - 4 \tau \i k_j^3}\right)\frac{2\pi k_j}{\beta\eps},
\end{equation}
and the jump in the remainder is therefore given by
\begin{equation}\label{A:2DSN}
\left[S_N\right]_-^+ \sim \left(\frac{1}{1 - 4 \tau \i k_j^3}\right)\frac{2\pi k_j \Theta}{\beta\eps}\e^{-k_j(w+\i\pi)/\eps}.
\end{equation}
The corresponding complex conjugate contribugion is switched across Stokes curves generated by the singularity at $w = \i\pi$ in the analytically continued free surface. Hence, the combined expression for the waves is given by twice the real part of \eqref{A:2DSN}.

\section{Detailed Analysis for Three-Dimensional Geometry}
\subsection{Prefactor Equation}\label{PREFACTOR_APP}

To find the prefactor equation, we consider the next order in \eqref{1:lbc1}--\eqref{1:lbc2} as $n \rightarrow \infty$. In order to uniquely determine the prefactors, we must expand $\Phi$ and $\Xi$ as power series in the limit that $n \rightarrow \infty$, and determine a consistency condition. We write
\begin{equation}
\Phi = \Phi_0 + \frac{1}{n}\Phi_1 + \ldots,\qquad \Xi = \Xi_0 + \frac{1}{n}\Xi_1 + \ldots.
\end{equation}
Applying the late-order ansatz to \eqref{1:sergeq}--\eqref{1:serbc2} now gives
\begin{align}\nonumber
-\chi_z\Phi_1 +\chi_x\Xi_1 =& - \Phi_{0,z} + \Xi_{0,x} ,\\ \nonumber
\chi_x\Phi_1 + (\chi_x^2 + \chi_y^2)^2\Xi_1  =&  \Phi_{0,x}+4(\chi_x^3 + \chi_x \chi_y^2)\Xi_{0,x} +4 (\chi_y^3 + \chi_x^2 \chi_y )\Xi_{0,y} \\&+ (8\chi_x\chi_y\chi_{xy}+(6\chi_x^2 + 2\chi_y^2)\chi_{xx} + (6\chi_y^2 + 2\chi_x^2)\chi_{yy})\Xi_0  .
\end{align}
This system only has nontrivial solutions for $\Phi_1$ and $\Xi_1$ when 
\begin{align}\nonumber
\chi_x(\Phi_{0,z} - \Xi_{0,x}) = \chi_z(&  \Phi_{0,x}+4(\chi_x^3 + \chi_x \chi_y^2)\Xi_{0,x} +4 (\chi_y^3 + \chi_x^2 \chi_y )\Xi_{0,y} \\&+ (8\chi_x\chi_y\chi_{xy}+(6\chi_x^2 + 2\chi_y^2)\chi_{xx} + (6\chi_y^2 + 2\chi_x^2)\chi_{yy})\Xi_0).
\end{align}
For ease of notation we now omit the subscripts and denote $\Xi_0$ by $\Xi$ and $\Phi_0$ by $\Phi$. This therefore gives
\begin{align}
\label{1:phiz} \Phi_z = \Xi_x + \frac{\chi_z}{\chi_x}&  \Phi_{x}+4(\chi_x^3 + \chi_x \chi_y^2)\Xi_{x} +4 (\chi_y^3 + \chi_x^2 \chi_y )\Xi_{y} \\&+ (8\chi_x\chi_y\chi_{xy}+(6\chi_x^2 + 2\chi_y^2)\chi_{xx} + (6\chi_y^2 + 2\chi_x^2)\chi_{yy})\Xi.\nonumber
\end{align}

In order to solve the prefactor equation \eqref{1:ansgov2}, we will express the equation on the free surface entirely in terms of $x$ and $y$ derivatives. This will result in an equation that has the exact same ray structure as the singulant equation \eqref{1:eik}, and hence the solution may be obtained in terms of the same characteristic variables. The equations from \eqref{1:phiz} give appropriate expressions for $\chi_z$ and $\Phi_z$; however, we must still consider the second derivative terms that will appear in the equation. Taking derivatives of $\chi_z$ and rearranging gives
\begin{align}
\label{1:chixz}\chi_{xz} &= -\frac{2\chi_x\chi_{xx}}{(\chi_x^2+\chi_y^2)^2} + \frac{4\chi_x^2(\chi_x\chi_{xx}+\chi_y\chi_{xy})}{(\chi_x^2+\chi_y^2)^3},\\
\label{1:chiyz}\chi_{yz} &= -\frac{2\chi_x\chi_{xy}}{(\chi_x^2+\chi_y^2)^2} + \frac{4\chi_x^2(\chi_x\chi_{xy}+\chi_y\chi_{yy})}{(\chi_x^2+\chi_y^2)^3},\\
\label{1:chizz}\chi_{zz} &= -\frac{2\chi_x\chi_{xz}}{(\chi_x^2+\chi_y^2)^2} + \frac{4\chi_x^2(\chi_x\chi_{xz}+\chi_y\chi_{yz})}{(\chi_x^2+\chi_y^2)^3}.
\end{align}
The final expression can be simplified using \eqref{1:chixz}--\eqref{1:chiyz} to completely eliminate the $z$-dependence from $\chi_{zz}$. Using \eqref{1:chixz}--\eqref{1:chizz}, as well as \eqref{1:phixi}, and \eqref{1:phiz} we write the prefactor equation \eqref{1:ansgov2} in terms of $x$ and $y$ derivatives on $z=0$ as
\begin{equation}
 \label{1:sspfeq}\left[4\chi_x^3 + 10\chi_x(\chi_x^2+\chi_y^2)^4\right]\Phi_x + \left[10\chi_y(\chi_x^2 + \chi_y^2)^4\right]\Phi_y = G(x,y)\Phi, 
\end{equation}
where
\begin{align}
G(x,y) = (1-6 \chi_x^2\chi_y^2(\chi_x^2 - \chi_y^2)) \chi_{xx} + 8 \chi_x^3 \chi_y (\chi_x^2 - 2 \chi_y^2) \chi_{xy} + (1 - 2\chi_x^6 + 10 \chi_x^4 \chi_y^2)\chi_{yy}.
\end{align}
This equation may be solved using the method of characteristics, giving the ray equations in terms of characteristic variable $u$ as
\begin{equation}
\label{1:ssrays2}\diff{x}{u} = 4\chi_x^3 + 10\chi_x(\chi_x^2+\chi_y^2)^4,\qquad \diff{y}{u} = 10\chi_y(\chi_x^2 + \chi_y^2)^4,\qquad \diff{\Phi}{u} = G(x,y)\Phi.
\end{equation}
The first two of these equations govern the ray paths, and are identical to the ray equations associated with \eqref{1:eik}. This allows \eqref{1:ssrays2} to be written in terms of the associated Charpit variables, and solved to give 
\begin{equation}
 \label{1:Phiss1} \Phi(s,u) = \Phi(s,0)\left[1 + \frac{10(s^6-2h^2s^4)u}{3h^7}\right]^{^{\frac{(-1)^{1/3} (4h^{10}-5h^8s^2-30h^2s^8-48s^{10})}{30 h^{8/3}s^{16/3}(2h^2-s^2)}}},
 \end{equation}
where the characteristic variable $u$ the same characteristic variable as the singulant, given by
\begin{align}
u = -\frac{h^7 (s - x)}{2 s^5 (2 h^2 + 5 s^2)}.
\end{align}
This provess can be systematically performed using standard computational algebra programming methods. Selecting the corresponding expression for $s$ in terms of $x$ and $y$ from \eqref{1:sspoly} gives the solution in terms of the physical coordinates $x$ and $y$. To find an expression for $\Phi(s,0)$, the behaviour of the system in the neighbourhood of $u = 0$ must be computed and matched to this outer solution. 

\subsection{Inner Expansion Near Singularity}\label{CH5_INNER1}

To solve the inner problem, we first consider the behaviour of $\chi$ near the singularity at $x^2 + y^2 + (z+h)^2 = 0$, which takes the form
\begin{equation}
\label{1:innerchi1} \chi_{L1} \sim \frac{x^{2/3}}{2h^{5/3}}\left(x^2 + y^2 + (z+h)^2\right).
\end{equation}
In the prefactor equation \eqref{1:Phiss}, we see that the unknown coefficient is a function of $s$. From the Charpit analysis, it follows that $s \sim x$ near the singularity at $t = 0$. Hence, we define a system of inner coordinates given by
\begin{equation}
\label{1:innercoords} \eps \sigma_1 = \frac{x^{2/3}}{2h^{5/3}}\left(x^2 + y^2 + (z+h)^2\right),\qquad\eps \sigma_2 = \frac{x^{2/3}}{2h^{5/3}}\left(x^2 + y^2 + (z-h)^2\right),\qquad\lambda = x.
\end{equation}
To leading order in $\eps$, the linearized governing equation \eqref{1:lgeq} becomes
\begin{equation}
 5\sigma_1\phi_{\sigma_1\sigma_1} + 5\sigma_2\phi_{\sigma_2\sigma_2} + \lambda\phi_{\lambda\sigma_2} + \lambda\phi_{\lambda\sigma_1} = 0,
\end{equation}
where terms containing derivatives with respect to both $\sigma_1$ and $\sigma_2$ were disregarded due to the form of the inner expansion, \eqref{1:localseries1}. Similarly, the boundary conditions \eqref{1:lbc1}--\eqref{1:lbc2} become 
\begin{alignat}{2}
\label{1:localbc1} h\phi_{\sigma_1} - h\phi_{\sigma_2} - \lambda\xi_{\sigma_1} -\lambda\xi_{\sigma_2}&= 0 \qquad & \mathrm{on} \quad \sigma_1 &= \sigma_2,\\
\label{1:localbc2} h\phi_{\sigma_1} +h\phi_{\sigma_2} -\lambda\xi_{\sigma_1\sigma_1\sigma_1\sigma_1} - \lambda\xi_{\sigma_2\sigma_2\sigma_2\sigma_2} &= 0  \qquad & \mathrm{on} \quad \sigma_1 &= \sigma_2.
\end{alignat}
Finally, by expressing the leading-order behaviour \eqref{1:phi0} in terms of the local variables, we find that
\begin{equation}
\label{1:localphi0} \phi^{(0)} \sim \frac{\lambda^{1/3}\sqrt{2}}{8\pi h^{5/6}\eps^{1/2}\sigma_1^{1/2}}-\frac{\lambda^{1/3}\sqrt{2}}{8\pi h^{5/6}\eps^{1/2}\sigma_2^{1/2}}.
\end{equation}
We now define the series expansion near the singularity on the complexified free surface as
\begin{align}
\label{1:localseries1} \phi \sim \sum_{n=0}^{\inf} \left[\frac{a_n(\lambda)}{\sigma_1^{n+1/2}} + \frac{b_n(\lambda)}{\sigma_2^{n+1/2}}\right], \qquad \xi \sim \sum_{n=0}^{\inf} \left[\frac{2c_n(\lambda)}{\sigma_1^{n+1/2}}\right],
\end{align}
where the latter expression is only valid on the free-surface itself, on which $\sigma_1 = \sigma_2$. The factor of two is included for subsequent algebraic convenience, and has no effect on the solution to the problem as $c_n$ is unknown at this stage of the analysis. From \eqref{1:localphi0}, we have
\begin{equation}
\label{1:locala0b0} a_0(\lambda) = \frac{\lambda^{1/3}\sqrt{2}}{8\pi h^{5/6}},\qquad b_0(\lambda) = -\frac{\lambda^{1/3}\sqrt{2}}{8\pi h^{5/6}}.
\end{equation}
We are interested in the behaviour of the terms on the complexified free surface in the neighbourhood of the singularity at $x^2 + y^2 + h^2 = 0$. Consequently, we apply the series expression to \eqref{1:localbc1} on the surface (defined by $\sigma_1 = \sigma_2$) and match in the limit that $\sigma_1$ (and therefore $\sigma_2$) tend to zero, giving
\begin{equation}
\label{1:localsurf1} -h(a_n - b_n) - 2\lambda c_n = 0, \qquad n \geq 0.
\end{equation}
Applying the series expansion to \eqref{1:localbc2} and matching in the same limit gives
\begin{equation}
\label{1:localsurf2} -h(n+7/2)(n+5/2)(n+3/2)(a_n + b_n) + 2 c_{n+1} = 0, \qquad n \geq 0.
\end{equation}
We are interested in the behaviour on the complexified free-surface; however, restricting the domain in this fashion means that it is impossible to distinguish between the contributions from the series in $\sigma_1$ and the series in $\sigma_2$. We note, however, that the two contributions have equal magnitude in \eqref{1:localphi0}. As the singular behaviour of the problem is preserved in all higher orders \cite{Dingle1}, we conclude that this must be true of the contributions at all subsequent orders. We therefore set $|a_n| = |b_n|$ in order to maintain consistency with the leading-order singularity contributions. This may only be accomplished if we divide the two equations given in \eqref{1:localsurf1}--\eqref{1:localsurf2} into four equations such that
\begin{alignat}{2}
 -h a_n- \lambda c_n &= 0, \qquad\qquad &-h (n+7/2)(n+5/2)(n+3/2) a_n+ \lambda c_{n+1} &= 0,\\
 h b_n- \lambda c_n  &= 0, \qquad\qquad &-h (n+7/2)(n+5/2)(n+3/2) b_n+ \lambda c_{n+1}&= 0.
\end{alignat}
We will consider only the first two of these equations, noting that the remaining equations imply that $b_n = (-1)^n a_n$. Eliminating $c_n$ from this system gives
\begin{equation}
 a_{n+1} = (n+7/2)(n+5/2)(n+3/2)a_n = \frac{a_0 \Gamma(3n+1/2)}{\Gamma(1/2)}.
\end{equation}
Hence, using the expression for $a_0$ given in \eqref{1:locala0b0}, we may match the local series expression given in \eqref{1:localseries1} with the prefactor given in \eqref{1:Phiss}. Noting that $\lambda$ is the local expression for $s$ in the outer solution, and that $\Phi(s,0)$ in the outer coordinates matches with $a_n(\lambda) + b_n(\lambda)$ in the inner coordinates, we find that
\begin{equation}
\label{1:arbitraryprefactor} \Phi(s,0) =  \frac{s^{1/3}\sqrt{2}}{4\pi^{3/2} h^{5/6}}.
\end{equation}
Hence, we are able to completely describe the late-order behaviour of terms in \eqref{1:series}, with the complete expression given in \eqref{1:Phiss}.

\subsection{Exponential Asymptotic Analysis}\label{CH5_STEADYSMOOTHING}

The asymptotic series given in \eqref{1:series} may be truncated to give
\begin{equation}
 \label{1:series2}\tilde{\phi} = \sum_{n=0}^{N-1}\eps^n\phi^{(n)} + R^{(N)},\qquad \tilde{\xi} = \sum_{n=0}^{N-1}\eps^n\xi^{(n)} + S^{(N)},
\end{equation}
where $N$ will be chosen in order to minimize the remainders $R^{(N)}$ and $S^{(N)}$. Applying this series expression to \eqref{1:lgeq} gives
\begin{equation}
 \label{1:ssge}\nabla^2 R^{(N)} = 0,
\end{equation}
while the boundary conditions \eqref{1:lbc1}--\eqref{1:lbc2} become on $z = 0$
\begin{align}
 \label{1:ssit0}  R^{(N)}_z - S^{(N)}_x &= 0,\\
 \label{1:ssit} R^{(N)}_x + \ \eps^3 \left( S^{(N)}_{xxxx} + 2 S^{(N)}_{xxyy} + S^{(N)}_{yyyy} \right)&= -\eps^{3N}(\xi^{(N-1)}_{xxxx} + 2\xi^{(N-1)}_{xxyy} + 2\xi^{(N-1)}_{yyyy}),
\end{align}
having made use of the relationship in \eqref{1:serbc2} and the fact that $\phi_x^{(0)} = 0$. The homogeneous form of \eqref{1:ssge}--\eqref{1:ssit} is satisfied as $\eps \rightarrow 0$ by
\begin{equation}
  \label{1:ssrt2}R^{(N)} \sim \Phi\e^{-\chi/\eps},\qquad S^{(N)} \sim \Xi\e^{-\chi/\eps},
\end{equation}
where $\chi$ is one of the singulants determined from \eqref{1:sschiray2}--\eqref{1:sspoly}. 

 We therefore set the remainder terms for the inhomogeneous problem to take the form
\begin{equation}
  \label{1:ssrt}R^{(N)} = A(x,y,z)\Phi\e^{-\chi/\eps},\qquad S^{(N)} = B(x,y)\Xi\e^{-\chi/\eps},
\end{equation}
where $A$ and $B$ are Stokes switching parameters. From \eqref{1:ssit0}, we see that $A = B$ on $z=0$. 

To determine the late order term behaviour, we will require the first correction term for the prefactors, and we therefore set
\begin{equation}
\Phi = \Phi_0 + \eps \Phi_1 + \ldots, \qquad \Xi = \Xi_0 + \eps \Xi_1 + \ldots.
\end{equation}
Applying the remainder forms given in \eqref{1:ssrt} to the boundary conditions , \eqref{1:ssit0} and \eqref{1:ssit}, gives after some rearrangement
\begin{align}\nonumber
-A\chi_z\Phi_1 +A\chi_x\Xi_1 =& - A\Phi_{0,z} - A_z \Phi_0+ A\Xi_{0,x} +A_x \Xi_0,\\
\nonumber A\chi_x\Phi_1 + A(\chi_x^2 + \chi_y^2)^2\Xi_1  =&  A \Phi_{0,x}+ 4B(\chi_x^3 + \chi_x \chi_y^2)\Xi_{0,x} +4 A (\chi_y^3 + \chi_x^2 \chi_y )\Xi_{0,y} \\
\nonumber& + A_x \Phi_{0}+ 4A_x(\chi_x^3 + \chi_x \chi_y^2)\Xi_{0} +4 A_y (\chi_y^3 + \chi_x^2 \chi_y )\Xi_{0}\\
\nonumber&+A (8\chi_x\chi_y\chi_{xy}+(6\chi_x^2 + 2\chi_y^2)\chi_{xx} + (6\chi_y^2 + 2\chi_x^2)\chi_{yy})\Xi_0\\
&+\eps^{3N-2}\e^{\chi/\epsilon}(\xi^{(N-1)}_{xxxx} + 2\xi^{(N-1)}_{xxyy} + 2\xi^{(N-1)}_{yyyy}).
\end{align}
Combining these expressions, and making use of \eqref{1:phiz} to eliminate terms and \eqref{1:serbc2} to simplify the right-hand side gives
\begin{equation}
-\chi_x(A_z\Phi_0 - A_x \Xi_0) + \chi_z(A_x \Phi_{0}+ 4(\chi_x^2 + \chi_y^2)(A_x\chi_x + A_y \chi_y)\Xi_{0})  \sim -\epsilon^{3N-2}\chi_z (\chi_x^2 + \chi_y^2)^2 \xi^{(N-1)} \e^{\chi/\eps}.
\end{equation}
As only the leading order prefactor behaviour appears in the final expression, we will no longer retain the subscripts. Applying the late-order ansatz gives
\begin{equation}
-\chi_x(A_z\Phi - A_x \Xi) + \chi_z(A_x \Phi+ 4(\chi_x^2 + \chi_y^2)(A_x\chi_x + A_y \chi_y)\Xi)  \sim  \eps^{3N-2} \frac{\chi_x^2\Xi\Gamma(3N-3/2)}{\chi^{3N-3/2}} \e^{\chi/\eps}.
\end{equation}
Motivated by the homogeneous solution, we express the equation in terms of $\chi$ and $y$, and apply \eqref{1:phixi} to obtain 
\begin{equation}
 3 A_{\chi}  = \eps^{3N-2}\e^{{\chi}/\eps}\frac{\Gamma(3N-3/2)}{{\chi}^{3N-3/2}}.
\end{equation}
The optimal truncation point is given by $N \sim |\chi|/3\eps$ in the limit that $\eps \rightarrow 0$. We write ${\chi} = r\e^{\i\theta}$, with $r$ and $\theta$ real so that $N = r/3\eps + \alpha$, where $\alpha$ is necessary to make $N$ an integer. Since $N$ depends on $r$ but not $\theta$, we write
\begin{equation}\label{CH2S1_TruncatedSeriesRescaling1}
 \pdiff{}{{\chi}} = -\frac{\i\e^{-\i\theta}}{r}\pdiff{}{\theta}.
\end{equation}
Using Stirling's formula on the resultant expression gives 
\begin{equation}
 A_{\theta}\sim\frac{\i \sqrt{2\pi r}}{3\eps}\exp\left(\frac{r}{\eps}\left(\e^{\i\theta}-1\right) - \i\theta\left(\frac{r}{\eps}+3\alpha-\frac{1}{2}\right)\right).
\end{equation}
This variation is exponentially small, except in the neighbourhood of the Stokes line, given by $\theta = 0$, where it is algebraically large. To investigate the rapid change in $A$ in the vicinity of the Stokes line, we set $\theta = \eps^{1/2}\hat{\theta}$, giving
\begin{equation}
 A_{\hat{\theta}} \sim {\frac{\i}{3} \sqrt{\frac{2\pi r }{\eps}}}\e^{-r\hat{\theta}^2/2},
\end{equation}
so that
\begin{equation}
 A\sim \frac{\i}{3}\sqrt{\frac{2\pi}{\eps}}\int_{-\inf}^{\theta \sqrt{r/\eps}} \e^{-t^2/2} \d t + C,
\end{equation}
where $C$ is constant. Thus, as the Stokes line is crossed, $A$ rapidly increases from 0 to $2\pi\i\eps^{-1/2}$. Using \eqref{1:ssrt}, we find the variation in the fluid potential, and we subsequently use \eqref{1:ssit0} to relate $B$ to $A$. We therefore find the variation in the free surface behaviour as the Stokes line is crossed. The Stokes line variation for the potential and free surface position are respectively given by
\begin{equation}
 \label{1:ssvariation} \left[R^{(N)}\right]_-^+ =\frac{2\pi\i\Phi}{3\sqrt{\eps}}\e^{-\chi/\eps},\qquad \left[S^{(N)}\right]_-^+ =\frac{2\pi\i\Xi}{3\sqrt{\eps}}\e^{-\chi/\eps},
\end{equation}
Hence, if we determine the prefactor and singulant behaviour associated with each contribution, \eqref{1:ssvariation} gives an expression for the behaviour switched on across the appropriate Stokes line. 

\bibliographystyle{plain}
\bibliography{sydrefs2.bib}

\end{document}